\documentclass[11pt, preprint]{aastex}
\usepackage{graphicx}
\begin{document}

\title{NUMERICAL SIMULATIONS OF THE ONSET AND STABILITY OF DYNAMICAL MASS TRANSFER
IN BINARIES}

\author{Mario C. R. D'Souza, Patrick M. Motl, Joel E. Tohline and Juhan Frank}
\affil{Department of Physics and Astronomy, Louisiana State
University,
    Baton Rouge, LA 70803}

\pagenumbering{arabic} \pagestyle{headings}

\begin{abstract}
Hydrodynamical simulations of semi-detached, polytropic binary stars
are presented in an effort to study the onset and stability of
dynamical mass transfer events. Initial, synchronously rotating
equilibrium models are constructed using a self-consistent-field
technique and then evolved with an Eulerian hydrodynamics code in a
fully self-consistent manner. We describe code improvements
introduced over the past few years that permit us to follow
dynamical mass-transfer events through more than 30 orbits.
Mass-transfer evolutions are presented for two different initial
configurations: A dynamically unstable binary with initial mass
ratio (donor/accretor) $q_0 = 1.3$ that leads to a complete merger
in $\sim 10$ orbits; and a double-degenerate binary with initial
mass ratio $q_0 = 0.5$ that, after some initial unstable growth of
mass transfer, tends to separate as the mass-transfer rate levels
off.
\end{abstract}

\keywords{Numerical Hydrodynamics --- Interacting binaries:
accretion --- Mass-transfer --- Stability --- Double White Dwarfs}

\section{Introduction}\label{Intro}

Most types of binary stars observed today in a long-term interacting
phase of their evolution --- such as cataclysmic variables, X-ray
binaries and contact binaries --- and other types that are no longer
actively transferring mass, are survivors of a dynamical phase of
evolution in a more violent past. Such rapid evolutionary phases
accompany the onset of mass transfer that occurs when angular
momentum losses or stellar evolutionary processes bring one
component of a detached system into contact with its Roche lobe. It
is important to investigate the onset and stability of the ensuing
mass transfer in order to understand the origin and the evolutionary
links between the various types of binary systems that we observe
today. The stability of a given binary system upon contact depends
on the system's mass ratio, the structure of the donor star, and how
mass and angular momentum is redistributed in the binary during the
ensuing mass-transfer event. By making reasonable assumptions that
encapsulate the essential physics while simplifying the governing
equations, analytical treatments can be devised that provide insight
into not only the stability, but the long-term evolutionary behavior
of mass-transferring, semi-detached binaries \citep{SoPh97, HaWe99,
GPF}. But because rapid mass-transfer events, in particular, are
intrinsically complex --- involving, for example, nonlinear
supersonic flows and a gravitational field that changes in a
nontrivial way in response to the dynamical exchange of matter
between the stars --- the rapid phases through which many binaries
evolve are not likely to be well understood until they are modeled
hydrodynamically with the inclusion of as much physics as possible.
In this paper, we demonstrate how large-scale numerical,
hydrodynamic simulations can be employed to better understand the
stability and the long-term evolutionary behavior of binaries as
they undergo rapid phases of mass-transfer. Note that in this paper
the term ``dynamical" is used to mean processes that result in
significant changes in the binary parameters and mass-transfer rates
in a few orbital periods.

Among several possible mechanisms driving binary evolution, close
binaries inevitably lose orbital energy and angular momentum by the
emission of gravitational waves. Double degenerate white dwarfs
(DWDs) are of particular interest since their expected numbers and
rates of emission of gravitational waves are such that they are
likely to constitute an important source of background noise for the
LISA mission \citep{HiBe00}. Details of the time-evolutionary
behavior of the gravitational-wave signal that is emitted from any
given system depend on the details of mass and angular momentum flow
in the binary. Reliable estimates for the rates of emission and the
time-frequency characteristics of the emitted signal (the
``template") depend on having a good understanding of the above
processes. While some progress can be made with analytical
prescriptions, theoretically constructed templates are not likely to
be useful for the extraction from the background of signals from
individual sources until detailed hydrodynamical models can
accurately simulate dynamical mass-transfer events in DWDs.
Furthermore, it has been realized recently that many of the
theoretically predicted and abundant detached DWDs that are driven
to contact in a Hubble time, would start mass transfer in a direct
impact mode \citep{Maet04}. In this mode, the accretion stream hits
the accretor's surface and dissipates without forming an accretion
disk. Therefore the orbital angular momentum carried by the stream
serves to spin up the accretor and is generally not returned by
tides to the orbit. One would like to study these processes without
the approximations usually adopted for mathematical convenience.
Since relatively little is known about these direct impact binaries,
it is also of interest to investigate the structure and properties
of these flows and explore possible observational signatures.

The above considerations have motivated us to undertake a long-term
program that utilizes detailed numerical hydrodynamic simulations to
model dynamical phases of mass transfer that can arise when binaries
of different types reach a semi-detached state. We have developed a
hydrodynamical code that is capable of following such binaries
through more than 30 orbits while conserving mass and angular
momentum to a high level of precision. As a first step, we studied
the evolution of detached polytropic binaries with components of
equal and unequal mass (Motl, Tohline \& Frank 2002, hereafter MTF).
In the present paper we describe some important improvements that
have been made to the code described by MTF and report on our first
results of simulations that follow dynamical mass-transfer events in
a fully self-consistent manner. We follow the evolution of two
different polytropic binary systems whose properties were selected
to illustrate the capabilities of our code and to provide some
overlap with earlier, related studies. In particular, one of our
models has been chosen specifically for comparison with a smooth
particle hydrodynamics (SPH) simulation that was presented a decade
ago by \citet{RS95}. To our knowledge, this work has not been
repeated with an Eulerian code in the intervening years.

Much of the related numerical simulation work that has preceded ours
(see \S\ref{PreviousSimulations} for elaboration) has concentrated
on stars having relatively stiff equations of state and on systems
that encounter and evolve through a tidal instability.  The
hydrodynamics of the coalescence, including brief phases of
mass-transfer, has been modeled when necessary as a prelude to an
almost inevitable merger. Our focus is, instead, on binaries whose
components obey a relatively soft equation of state.  Our objective
is to investigate in more detail the hydrodynamics of mass transfer,
the structures arising from this transfer, and the role mass
transfer plays during the dynamical phase, sometimes leading the
binary closer to a tidal instability and merger, but sometimes
saving it from merger and leading it to a slower evolutionary phase.

\section{Theoretical Background}\label{secBackground}

Our present investigation focuses on the dynamical stability of
semi-detached, polytropic binary star systems in circular (or very
nearly circular) orbits.  A polytrope is a star whose equation of
state is governed by the relation,
\begin{eqnarray}
p &=& K\rho^{1+1/n} \, ,
\end{eqnarray}
where $p$ is pressure, $\rho$ is mass density, $K$ is a constant
that defines the specific entropy of the gas, and $n$ is the
so-called polytropic index that specifies the degree of
compressibility of the gas. Each initial configuration is uniquely
defined by specifying: the mass of each star, $M_{\rm d}$ (the
donor) and $M_{\rm a}$ (the accretor); the orbital separation $a$
({\it i.e.}, the distance between the centers of mass of the two
stars); the polytropic index $n$; and each star's polytropic
constant, $K_{\rm d}$ and $K_{\rm a}$. Throughout this presentation,
we also will frequently refer to the system mass ratio,
\begin{eqnarray}
q \equiv \frac{M_{\rm d}}{M_{\rm a}} \, . \label{massRatioDefined}
\end{eqnarray}
Although our tools are capable of evolving binaries with any
polytropic index and even more general equations of state, here we
will only be considering polytropic binaries in which $n=3/2$. This
particular choice is of interest because key dynamical properties of
mass-transferring binary systems that contain low-mass main-sequence
(MS) stars or white dwarfs (WD) can be realistically modeled using
an $n=3/2$ polytropic equation of state, if an appropriate choice is
made for the polytropic constants of the two binary components.

For spherically symmetric polytropic stars, the radius of the star
$R$ is uniquely determined once the three parameters $n$, $M$, and
$K$ have been specified (cf., Chandrasekhar 1958).  For a given
polytropic index, the star's mass-radius relationship is uniquely
defined as well; specifically,
\begin{eqnarray}\label{PolytropicM_R}
M &=& k_n R^{(3-n)/(1-n)} \, ,
\end{eqnarray}
where the proportionality constant $k_n$ depends on $K$ and the
gravitational constant $G$ through an expression that
is determined from a solution of the relevant Lane-Emden equation
(cf., Chandrasekhar 1958), for example, $k_{3/2} = 13.1 [K/G]^{3}$.
In close binaries, both stars generally are rotationally flattened
and tidally distorted.  Hence, their geometric shape cannot be
accurately characterized by a single radius. Nevertheless, an
``effective'' radius $R_{\rm d}$ and $R_{\rm a}$ of the donor and
accretor, respectively, can be defined as the radius of a sphere
having a volume $V_{\rm d}$ or $V_{\rm a}$ that is filled by each
distorted star. The proportionality constant that relates the mass
to the effective radius is different from the value of $k_n$ that
holds for spherical polytropes, but the power-law in the mass-radius
relationship is approximately the same. For example, if the
structure of the donor is well described by a polytrope of index
$n=3/2$, the donor's effective radius increases as it loses mass
according to the relation,
\begin{eqnarray}
R_{\rm d} \propto M_{\rm d}^{-1/3} \, .
\label{MassRadiusRelationship}
\end{eqnarray}

\subsection{Concepts derived from the point-mass approximation}

For reference we summarize here, using the notation introduced in
this paper, some well-known results concerning mass transfer in
binaries. For more details, the interested reader may refer to
reviews by \citet{Pac71, RaJoWe82, HWe87, VeRa88, Ki88, SoPh97,
HaWe99, FKR}. If, for the moment, we assume that the stars are point
masses in a circular orbit, then three parameters ($M_{\rm d}$,
$M_{\rm a}$, and $a$) are sufficient to uniquely define the binary
system's orbital frequency,
\begin{eqnarray}
\Omega &=& \biggl[\frac{G(M_{\rm a}+M_{\rm d})}{a^3}\biggr]^{1/2} \,
,
\end{eqnarray}
and its total, purely orbital, angular momentum,
\begin{eqnarray}
J_\mathrm{tot}=J_\mathrm{orb} &=& \frac{M_{\rm d} M_{\rm a}}{M_{\rm
a}+M_{\rm d}}~ a^2\Omega = M_{\rm d} M_{\rm a} \biggl(\frac{G
a}{M_{\rm a}+M_{\rm d}}\biggr)^{1/2} \, . \label{jorb}
\end{eqnarray}
For a given total mass, $M_\mathrm{tot} = (M_a + M_d)$, and angular
momentum, $J_\mathrm{tot}$, then, it is easy to show that the binary
system will have its minimum separation when $M_{\rm d} = M_{\rm
a}$, that is, when $q=1$. More generally, in a binary whose orbital
evolution is driven by systemic angular momentum losses
$(\dot{J})_\mathrm{sys}$ such as gravitational radiation, the
separation will evolve according to
\begin{equation}
\frac{\dot a}{2a} = \left(\frac{\dot J}{J_{\rm orb}}\right)_{\rm
sys} - \frac{\dot M_{\rm d}}{M_{\rm d}}\left(1-q\right)\, ,
\label{adotpoint}
\end{equation}
where the dots indicate differentiation with respect to time, and we
have assumed $\dot M_{\rm a}= -\dot M_{\rm d}$, so that mass is
conserved.

For a point-mass binary, the critical Roche surface can be defined
analytically in implicit form ({\it e.g.}, Frank et al. 2002).
The effective radius $R_L$ of the Roche lobe around the donor is
also well-defined.  As shown by \citet{Eggleton83}, the ratio
$R_L/a$ is only a function of the mass ratio, $q$, and it is fairly
accurately given by the approximate expression,
\begin{eqnarray}
\frac{R_L}{a} \approx \frac{0.49 q^{2/3}}{0.69q^{2/3} +
\ln(1+q^{1/3})} \, . \label{Eggleton}
\end{eqnarray}
A simpler to use, but somewhat cruder approximation, correct to
within 6\% in the range $0<q<4$ is due to Pacy\'nski (1971):
\begin{eqnarray}
\frac{R_L}{a} \approx 0.4622 \left(\frac{q}{1+q}\right)^{1/3} \, .
\label{Pacynski}
\end{eqnarray}
With these ideas in mind, if angular momentum and mass are conserved
during a mass-transfer event (fully conservative mass transfer), the
binary separation is smallest when $q=1$, and the Roche lobe is
smallest when $q\approx 5/6$. When mass transfer occurs in binaries
where the donor may be approximated as an $n = 3/2$ polytrope, the
donor expands when it loses mass according to the radius-mass
relationship given by expression (\ref{MassRadiusRelationship}). The
rate of mass transfer in a semi-detached binary in which the donor
slightly overfills its Roche lobe depends mainly on the depth of
contact which is proportional to $R_{\rm d}-R_L$. Clearly, this rate
increases or decreases according to whether the depth of contact
itself increases or decreases. With the above assumptions, it is
easy to show that
\begin{eqnarray}
\frac{\dot{R}_{\rm d}-\dot{R}_L}{R_{\rm d}} \approx
\frac{\dot{R}_{\rm d}}{R_{\rm d}}-\frac{\dot{R}_L}{R_L}=
\frac{-2\dot{M}_{\rm d}}{M_{\rm d}}(q-{2\over3}) \, .
\label{LinearStability}
\end{eqnarray}
Since $\dot{M}_{\rm d}$ is negative, the depth of contact increases
upon fully conservative mass transfer if
$q>q_{\mathrm{stable}}=2/3$, or decreases if
$q<q_{\mathrm{stable}}=2/3$. For a more complete discussion of the
stability of mass transfer in double-degenerate binaries see
\citet{HaWe99} (see also references cited therein). It therefore
proves instructive to divide our discussion of stability into two
broad regimes: $q>2/3$ and $q<2/3$.

\subsection{Expectations}\label{SecExpectations}

\subsubsection{$q>2/3$}

For $q > 2/3$, the donor expands more rapidly than the Roche lobe,
which may actually contract if $q>5/6$. The binary separation itself
expands if $q<1$ or contracts if $q>1$. Thus, if the donor is
initially more massive than the accretor, the orbital separation
will decrease with time and the effective radius of the Roche lobe
will decrease as well. Since the donor expands upon mass loss, mass
transfer will be clearly unstable as the Roche lobe encroaches on
the donor, even in the absence of driving. On the other hand, for
systems with $q_{\mathrm{stable}} < q \leq 5/6$, the donor expands
faster upon mass loss than its Roche lobe can expand and the mass
transfer rate will still grow with time. Of these systems, if $q$ is
initially only slightly above $q_{\mathrm{stable}}$, unstable mass
transfer may proceed until the mass ratio falls below the stability
limit. What happens thereafter depends on whether or not the system
falls prey to tidal instabilities. If it survives, the mass transfer
decays in the absence of driving or evolves toward stable mass
transfer if steady driving is present. Systems with an initial mass
ratio significantly higher than $q_\mathrm{stable}$ are likely to be
unstable and merge through a common envelope phase or the donor may
be tidally disrupted.

\subsubsection{ $q<2/3$ }

When $q < 2/3$, the donor is initially less massive than the
accretor. Hence, in the point-mass approximation as mass is
transferred from the donor to the accretor, the orbital separation
will increase with time, and $R_L$ will increase as well since
$q<5/6$. This will tend to stabilize the system because the Roche
lobe expands away from the original surface of the donor. Even as
the donor expands, in the absence of driving the depth of contact
will decrease and so will the mass-transfer. Thus in the absence of
driving, the mass transfer would ultimately decay to zero, while it
would tend to a stable value if driving is present.

Our discussion above assumes that the total angular momentum is
purely orbital and that mass transfer is fully conservative. This is
adequate provided that the accretion stream has sufficient angular
momentum to form a disk around the accretor. In that case tidal
torques on the disk will return the  angular momentum advected by
the stream back to the orbit. As discussed further below, even if
mass transfer is fully conservative, when the mass transfer stream
directly impacts the accretor and no disk forms, the advected
angular momentum spins up the accretor at the expense of the orbital
angular momentum reducing $q_{\mathrm{stable}}$ to values below 2/3.

\subsubsection{Direct Impact and Other Finite-Size
Effects}\label{SecDirectImpact}

When the finite size of the stars is taken into account, the total
angular momentum must be written as the sum of the orbital angular
momentum, $J_{\rm orb}$ as given by Eq.~(\ref{jorb}), and the spin
angular momenta of the donor $J_{\rm d}$ and the accretor $J_{\rm
a}$. Angular momentum may be exchanged between the orbit and the
binary components by advection and by tides. We cite here without
proof an equation that describes in an approximate fashion the rate
of change of the binary separation under the effects of a systemic
loss, direct impact and possible spin evolution due to tidal
interactions (see Marsh et al. 2004 or Gokhale et al. 2005 for
details):
\begin{equation}
\frac{\dot a}{2a} = \left(\frac{\dot J}{J_{\rm orb}}\right)_{\rm
sys}- \left(\frac{\dot J_{\rm a} + \dot J_{\rm d}}{J_{\rm
orb}}\right)_{\rm tides} - \frac{\dot M_{\rm d}}{M_{\rm
d}}\left(1-q-\sqrt{(1+q)r_h}\right)\, . \label{adot}
\end{equation}
Following \citet{Maet04}, the specific angular momentum that is
carried by the stream has been expressed in terms of the
circularization radius $r_h \equiv R_{\rm circ}/a$, and the term on
the right-hand-side of this equation that contains $r_h$ accounts
for the rate at which angular momentum is transferred by the stream
to the accretor.  The second term on the right-hand-side represents
the effects of purely tidal changes in the spin angular momenta.

As has been emphasized by \citet{MaSt02} and \citet{Maet04}, the
consequential loss of orbital angular momentum that accompanies
direct impact accretion (compare the last term on the
right-hand-sides of Eqs. \ref{adotpoint} and \ref{adot}) acts to
destabilize mass transfer. Recent semi-analytic work suggests that
this effect alone can reduce the stability limit from $2/3$ to a
value $q_\mathrm{stable} \approx 0.22$ \citep{GPF}. However, at the
onset of mass transfer in a DWD
--- especially if, initially, $q \ge q_{\mathrm{stable}}$ --- the
mass accretion rate $\dot{M}_a$ may well exceed the critical rate
that yields the Eddington luminosity and the excess mass may be
blown away. This effect will act to slightly increase
$q_{\mathrm{stable}}$ \citep{HaWe99}.

In this work, we relax the assumptions required to treat the
mass-transfer semi-analytically and instead investigate the
stability of $n=3/2$ polytropic binaries through direct
hydrodynamical simulations. Thus we are able to follow the internal
flow of mass and angular momentum via the stream and tides without
any of the assumptions usually adopted for mathematical convenience.
It should be kept in mind, however, that we do not include in this
investigation a self-consistent treatment of thermal relaxation, and
the effects of radiative transfer are ignored.

\subsection{Previous, Related, Hydrodynamical Simulations}
\label{PreviousSimulations}

The equilibrium and stability of polytropic binary sequences in
nearly circular orbits, including the synchronous and irrotational
cases, has been comprehensively discussed in a series of papers by
Lai, Rasio \& Shapiro \citep{LRS1, LRS2, LRS3, LRS4, LRS5} for
systems having various polytropic indexes and mass ratios. Their
results were confirmed and extended in a series of papers by
Rasio \& Shapiro (1992, 1994, 1995 --- henceforth RS92, RS94, RS95
respectively).
Using a relaxation method they constructed synchronously rotating
equilibrium binaries for various polytropic indexes, mass ratios and
initial separations, and followed their hydrodynamic evolution using
smoothed-particle hydrodynamics (SPH). In most cases the evolution
led to coalescence of the binary, although in one case that was
meant to represent binary neutron stars with a stiff equation of
state ($n=1/2$, $q= q_\mathrm{RS} = 0.5$) the model binary returned
to a new stable configuration after a phase of mass transfer (RS94).
Of particular interest to us in the context of this paper are the
results of RS95 who investigated the equilibrium and stability
properties of binaries with a variety of initial mass
ratios\footnote{In RS95, the parameter $q$ was defined as the ratio
of the less massive star to the more massive star so that $q\leq 1$
in all cases.  In order to avoid confusion, we will use
$q_\mathrm{RS}$ when referring to the mass-ratios quoted in RS95
then, in order to be consistent with the definition given here in
Eq.~(\ref{massRatioDefined}), we will set $q = 1/q_\mathrm{RS}$ when
the donor is initially more massive than the accretor.} and with
polytropic index $n = 3/2$.

For equal-mass binaries, both MS and WD binaries were represented in
RS95 by setting the polytropic constants of the two components to be
equal, that is, $K_{\rm d} = K_{\rm a}$. A sequence of equilibrium
configurations was constructed for a range of separations, $a$,
specified by the parameter $r \equiv a/R_{1}$, where $R_{1}$ was the
unperturbed radius of the more massive star (primary). These $q=
q_\mathrm{RS} = 1$, equilibrium binaries were then evolved in time
using the SPH method and systems with $r\lesssim 2.45$ were found to
suffer a dynamical instability. For MS binaries with $q_\mathrm{RS}
< 1$ (polytropic constants were adjusted so as to obtain the MS
mass-radius relation) it is the more massive star that overflows its
Roche lobe first so, consistent with the expectations discussed
above, RS95 found that the resulting mass transfer tended to be
unstable. Systems with $q_\mathrm{RS} \lesssim 0.4$ (that is, $q =
1/q_\mathrm{RS} \gtrsim 2.5$) were found by RS95 to be secularly
unstable even before the primary star filled its Roche lobe,
while a dynamical instability was encountered before the Roche limit
in binaries with  $q_\mathrm{RS} \lesssim 0.25$ (that is, $q =
1/q_\mathrm{RS} \gtrsim 4$). Finally, $q < 1$ systems with WD
components (the polytropic constants are set to be equal in order to
realize the appropriate WD mass-radius relationship, given here by
expression \ref{MassRadiusRelationship}) were found by RS95 to
remain secularly and dynamically stable until $r$ was small enough
for the donor to overflow its Roche lobe. The WD binary, $q = 0.5$,
was found to be unstable to mass transfer and the SPH simulation led
to tidal disruption of the donor star and final merger after five
orbital periods.

\subsection{In the Context of this Paper}
In \S\S\ref{sec_0.843_UB} and \ref{sec_mdot_results} of this paper,
we present the results of nine separate nonlinear hydrodynamical
simulations that we have conducted in an effort to better understand
mass-transfer instabilities in close binary systems and, at the same
time, to better understand the capabilities and limitations of our
numerical tools.  The initial models for these simulations were all
unequal-mass, synchronously rotating, $n=3/2$ polytropic binaries in
circular orbit. As is illustrated in Figure \ref{equipotentials}, we
have examined systems having three different initial mass ratios:
$q_0=0.843$, $q_0 = 1.323$, and $q_0= 0.500$.  The first of these
($q_0=0.843$) was designed to provide a comparison with the
benchmark ``model UB'' evolution that was discussed by MTF. The
initial model with $q_0=1.323$ (that is, $q_\mathrm{RS} = 1/q
\approx 0.76$) was designed to represent a MS binary in which the
more massive star makes first contact with the critical Roche
surface. The model with $q_0 = q_\mathrm{RS} = 0.5$ was designed to
represent a low-mass, DWD binary in which the less massive star has
the larger radius and therefore makes first contact with its Roche
lobe. Results from these new simulations will be compared with the
expectations that have been drawn from earlier analytical and
semi-analytical investigations, as well as with the nonlinear
simulations presented by RS95.  A particularly detailed comparison
will be made of the $q_0=0.5$ WD binary system since the parameters
defining the initial state of our model were chosen to match as
closely as possible the initial state that was investigated by RS95.

\section{Overview of the Computational Tools}\label{code_review}
The computational tools used in our present study are the same as
those used in MTF except for two modifications, which we describe in
detail in \S\ref{sec_corr_code}, below. In summary, we employ: A
self-consistent-field (SCF) code to construct each initial binary
model; and a three-dimensional, finite-difference hydrodynamics code
to evolve each initial model forward in time. Both codes utilize a
cylindrical computational grid with $R, \phi$ and $z$ denoting the
radial, azimuthal and vertical coordinates, respectively. As stated
earlier, our present study is confined to initial states in which
the fluid in both stars obeys an $n=3/2$ polytropic equation of
state, but in general $K_{\rm d} \neq K_{\rm a}$. During the
hydrodynamical evolutions, the state variables of every fluid
element vary in such a way that they follow adiabats having a ratio
of specific heats, $\gamma = 1+1/n = 5/3$.

Our SCF method is based on the iterative technique developed by
Hachisu (1986; see also Hachisu et al. 1986). In the past, this
technique has been used to construct initial conditions for
hydrodynamical studies of the relative stability of equal-mass
binary systems \citep{New97, SWC}. It also has been used to
construct a wide variety of unequal-mass, detached and semi-detached
binaries (MTF). Here we use the SCF code to build models of
synchronously rotating, unequal-mass binaries in a frame that is
corotating with the binary's initial orbital frequency $\Omega_0$,
so the two stars are stationary in this frame. The axis of rotation
is taken to be parallel to the $z$-axis of the coordinate grid. As
input to the code, we must specify: three boundary points on the
stars where the mass-density, $\rho$, must vanish; the maximum
density $\rho^\mathrm{max}$ of each star; and an initial guess for
the density distribution, $\rho(R,\phi,z)$. The three boundary
points lie along the line that joins the centers of the two stars;
they correspond to the inner and outer edges of one star (usually
the accretor), and the inner edge for the companion star. The SCF
code then iteratively solves for the initial equilibrium
configuration through the following steps. For the given density
distribution, the resulting Newtonian gravitational potential,
$\Phi(R,\phi,z)$, is calculated by solving Poisson's equation
(\ref{PoissonEq}). Using the value of $\Phi$ at the three boundary
points, the code determines the angular frequency of the binary
orbit and two integration constants. With these data, the enthalpy
$H(R,\phi,z)$ is computed within each star. Using $H$ and the
prescribed polytropic equation of state, an improved ``guess'' for
the density distribution is calculated. The iteration repeats until
a prescribed convergence criterion is met.

The hydrodynamical code has been designed to solve the equations
that govern the flow of inviscid, compressible, self-gravitating
fluids in a frame of reference that is rotating with an arbitrarily
chosen angular velocity (including zero, in which case it would be
an inertial reference frame).  Throughout this paper all the
simulations have been conducted in a frame that is rotating with the
binary system's {\em initial} orbital frequency, $\Omega_0$, as
derived from the SCF code. The primary variables that are evolved
forward in time by the hydrodynamical code are the volume-densities
of five conserved quantities: the mass density $\rho$, the radial
momentum density, $S$, the vertical momentum density, $T$, the
angular momentum density, $A$, and an entropy tracer, $\tau$. These
quantities are advanced in time via a conservative formulation and
an explicit integration of five, first-order hyperbolic partial
differential equations (\ref{EulerEqns1})-(\ref{entropyTracerEq}):
the three components of Euler's equation, and two continuity
equations (one for the mass density and one for the entropy tracer).
As described in \S 4 of MTF, the hydrodynamical time loop consists
of applying the source, advection and artificial viscosity operators
in a sequence that ensures a nearly second-order accurate
time-integration. This is in addition to enforcing the boundary
conditions and solving Poisson's equation for the gravitational
potential so that the fluid is accelerated in a self-consistent
Newtonian gravitational field.

In mass-transfer systems, flow across the L1 Lagrange point is
generally expected to be transonic and, thereafter, the
mass-transfer stream can quickly acquire Mach numbers $\gtrsim 10$
as it falls toward the accretor.  It is therefore not surprising
that in the mass-transfer simulations presented here, relatively
strong shocks develop as the accretion stream obliquely impacts the
surface of the accretor.  With this in mind, it is worth reviewing
how shocks are handled in our hydrocode (see MTF for more details).
In the vicinity of a shock, our (normally) second-order-accurate
advection scheme (using Van Leer monotonic interpolation) is reduced
to a first-order-accurate scheme to ensure numerical stability, and
artificial viscosity is introduced to mediate the shock, spreading
the (ideally, infinitesimally thin) shock front over a small number
of computational grid zones.  (The extra ``artificial viscosity''
source terms that are introduced into the three components of
Euler's equation to accomplish this task are not shown in our \S3.2
summary of the equations, but they are enumerated in MTF.)  As a
result, momentum is properly conserved across all shocks.  Mass is
also properly conserved, as the equation of continuity (Eq.~18)
remains unchanged in the presence of shocks.

Finally, energy will be conserved in an adiabatic shock only if the
specific entropy of material increases as it moves through the
shock.  We have not added source terms to the energy equation in our
hydrocode (Eq.~19) to account for this generation of entropy.  As a
result, material in the accretion stream retains its ``pre-shock''
specific entropy (it remains on precisely the same $\gamma = 5/3$
adiabat as it moves from one side of the shock to the other); and
post-shock densities and pressures are somewhat higher than would be
expected if energy were conserved and the so-called ``adiabatic
shock jump conditions'' \citep{LaLi} were
realized.  With this implementation of the energy equation, we are
effectively assuming that the post-shock gas instantaneously
radiates away the ``extra'' heat that should have been generated by
the shock. While one might argue that it is unreasonable to expect
radiation to cool the gas back down to precisely its pre-shock
entropy condition, it also seems unreasonable to expect that
material immediately behind a realistic accretion shock will not be
subject to some amount of radiative cooling.  Our handling of the
energy equation in the presence of a shock effectively provides a
measure of cooling, and does so in a manner that is straightforward
to implement numerically and readily reproducible by others who
might choose to use our simulations as a benchmark for further work.

\subsection{Recent Modifications to the Hydrodynamical Code}\label{sec_corr_code}

The task of self-consistently modeling the dynamical evolution of a
binary system that is undergoing mass transfer is computationally
challenging, as can be understood from the following considerations.
The vast majority of material in the hydrodynamical simulation is
gravitationally confined within the two stars and is nearly at rest
(as viewed from the rotating reference frame).  Mild deviations from
this state occur in response to the exchange of mass and momentum
between the two stars on a timescale that is on the order of the orbital
period.  At the same time, the dynamics of the mass-accretion stream
--- which generally will contain supersonic flows that are confined
to a relatively small volume of the computational domain
--- must be accurately resolved and may severely limit the size of
the time step that is permitted by the explicit integration scheme.
The evolutionary code must maintain the approximate force balance
within and between both stars to a high degree of precision in order
to permit an accurate treatment of the response of the binary to
Roche-lobe overflow.  In this context, it is important to emphasize
that, as is traditional in the astrophysical fluid dynamics
community, our hydrodynamics code has been developed around a
conservative formulation of the dynamical equations. This, in
itself, ensures that the advection operators preserve the integral
of conserved densities to machine precision. However, we are solving
a more complex problem where the fluid flow is coupled to the
gravitational field through Poisson's equation and, in the presence
of the related gravitational source terms, the code, in its
entirety, is no longer strictly conservative.

Two modifications to the algorithm described by MTF have been
crucial to the recent success of our Roche-lobe overflow
simulations. As is detailed in the following two subsections, both
have involved subtle modifications to the source terms in Euler's
equation.

\subsubsection{Treatment of Pressure Gradients}
As has been described by MTF, the inertial-frame source terms in
Euler's equation were originally written as the gradient of an
effective potential that includes the fluid enthalpy.  This
formulation of the source terms is consistent with the manner in
which initial equilibrium structures are constructed in the SCF
code. However, when fluid that is tidally stripped from the donor
falls directly onto the surface of the accretor, this formulation
produces incorrect pressure gradients if the material from the donor
has a different specific entropy from the material in the accretor.
Accordingly, we have modified the source terms in Euler's equation
to properly account for gradients in the fluid's specific entropy.
The hydrocode now explicitly calculates ${\bf \nabla}p$ instead of
the product $\rho{\bf \nabla}H$, as was indicated in the MTF code
description. Implications of this change are enumerated in later
sections of this paper.


\subsubsection{Center-of-Mass Motion}
\label{commotion}

By following the evolution of an unequal-mass, detached binary
system (model ``UB'') through just over five orbits, MTF showed that
the code conserves angular momentum to an accuracy $\Delta J_z/J_z
\approx 10^{-4}$ per orbit.  (See the first row of our Table 3 for a
summary of related measurements reported by MTF for model UB.)  They
also showed that, because linear momentum was not precisely
conserved, the center of mass of the binary system slowly wandered
away from its initial position (the center of the cylindrical
coordinate grid).  As Figure 15 of MTF illustrates, by the end of
the UB model evolution (just over five orbits), the center of mass
of the system had moved a distance $\approx \Delta R/4$ away from
the center, where $\Delta R$ was the radial size of one grid zone.
As MTF pointed out, this level of momentum conservation is excellent
when compared to the results of other groups who have performed
simulations of comparable (or even less) complexity using
finite-difference hydrodynamical schemes.

As we embarked upon our present project to model mass transfer in
semi-detached binary systems, we were concerned that even very slow
motion of the center of mass away from the center of the
computational grid might cause problems in evolutions that were
followed through significantly more than five orbits.  Most
importantly, we were concerned that the surface of one or both stars
might run into the outer edge of the computational grid as the
center of mass of the system wandered farther and farther from the
center of the grid. In an effort to confine the center of mass of
the system to a region very close to the center of the coordinate
grid, we have added a small ``artificial'' acceleration,
\begin{eqnarray}
{\bf a}^\mathrm{art} = {\bf e}_R a_R^\mathrm{art}+{\bf e}_\phi
a_\phi^\mathrm{art}+{\bf e}_z a_z^\mathrm{art}
\end{eqnarray}
to the source terms of Euler's equation that is designed to
counteract the empirically measured rate at which the center of mass
was otherwise being accelerated.

As viewed from an inertial reference frame, it is easy to explain
how the requisite size and direction of this artificial acceleration
vector should be estimated. Simply ``measure'' the size and
direction of the residual (and unphysical) center-of-mass
acceleration ${\bf [\ddot{r}}_\mathrm{com}]_\mathrm{inertial}$ at
any point in time, then set ${\bf a}^\mathrm{art} = - [\ddot{\bf
r}_\mathrm{com}]_\mathrm{inertial}$. Because our simulations have
been performed in a frame of reference that is rotating with a
constant angular velocity ${\bf \Omega} = {\bf e}_z \Omega_0$, we
have used the standard reference frame transformation,
\begin{eqnarray}
[\ddot{\bf r}]_\mathrm{inertial} &=& [\ddot{\bf r} + 2{\bf
\Omega\times \dot{r}} + {\bf \Omega\times}({\bf \Omega\times
r})]_\mathrm{rotating} \, ,
\end{eqnarray}
and determined ${\bf a}^\mathrm{art}$ from the expression,
\begin{eqnarray}
{\bf a}^\mathrm{art} &=& -~[\ddot{\bf r}_\mathrm{com} + 2{\bf
\Omega\times \dot{r}}_\mathrm{com} + {\bf \Omega\times}({\bf
\Omega\times r}_\mathrm{com})]_\mathrm{rotating} \, \nonumber \\
&=& - ~{\bf e}_R[a_{R,\mathrm{com}} - 2\Omega_0
v_{\phi,\mathrm{com}} - \Omega_0^2 R_\mathrm{com}] -~{\bf e}_\phi
[a_{\phi,\mathrm{com}} + 2\Omega_0 v_{R,\mathrm{com}}] -~{\bf e}_z
[a_{z,\mathrm{com}}] \, , \label{comAcceleration}
\end{eqnarray}
where it is understood in the last expression that all components of
the center-of-mass position, velocity, and acceleration are measured
in the rotating frame. Appendix \ref{comAppendix} details how, in
practice, each of the terms in Eq.~(\ref{comAcceleration}) is
evaluated in the hydrodynamic code.

We emphasize that, at each time step, the same vector ${\bf
a}^\mathrm{art}$ was included in the calculation of the acceleration
of every fluid element, but a new value of ${\bf a}^\mathrm{art}$
was determined every integration time step. Although this was not a
particularly sophisticated way to correct for center-of-mass motions
in our simulations, it proved to be quite successful. For example,
Figure \ref{Q0.5 xyRcom} shows that the motion of the center of mass
of a binary with $q_0=0.5$ was confined to within approximately one radial zone of
the center of the grid for over 30 orbits for the longest of the
three runs presented.

\subsection{Summary of Equations}
\label{equations} In contrast to the form of the three components of
Euler's equation that was presented in MTF, our present simulations
of mass-transferring binary star systems have utilized the following
equations:
\begin{eqnarray}
\frac{\partial S}{\partial t} + \mbox{\boldmath$\nabla$} \cdot (S
\mbox{\boldmath$\upsilon$} ) &=& - \frac{\partial p}{\partial R} -
\rho \frac{\partial}{\partial R} \left[ \Phi - \frac{1}{2}
\Omega_0^{2} R^{2} \right] + \frac{A^{2}}{\rho R^{3}} + 2 \Omega_0
\frac{A}{R} \nonumber \\
&& -\rho [a_{R,\mathrm{com}} -2\Omega_0 v_{\phi,\mathrm{com}}
-\Omega_0^2 R_\mathrm{com}] \, , \label{EulerEqns1} \\
\frac{\partial T}{\partial t} + \mbox{\boldmath$\nabla$} \cdot (T
\mbox{\boldmath$\upsilon$} ) &=& - \frac{\partial p}{\partial z} -
\rho \frac{\partial \Phi}{\partial z} -\rho[a_{z,\mathrm{com}}] \, ,
\label{EulerEqns2} \\
\frac{\partial A}{\partial t} + \mbox{\boldmath$\nabla$} \cdot (A
\mbox{\boldmath$\upsilon$} ) &=& - \frac{\partial p}{\partial \phi}
- \rho \frac{\partial \Phi}{\partial \phi} - 2 \Omega_0 S R - \rho
R[a_{\phi,\mathrm{com}} +2\Omega_0 v_{R,\mathrm{com}}] \, .
\label{EulerEqns3}
\end{eqnarray}
The statements of mass and entropy conservation, and the Poisson
equation remain as presented by MTF, namely,
\begin{eqnarray}
\frac{\partial \rho}{\partial t} + \mbox{\boldmath$\nabla$} \cdot
(\rho \mbox{\boldmath$\upsilon$} ) &=& 0 \, , \label{continuityEq} \\
\frac{\partial \tau}{\partial t} + \mbox{\boldmath$\nabla$} \cdot
(\tau \mbox{\boldmath$\upsilon$} ) &=& 0 \, , \label{entropyTracerEq} \\
\nabla^2\Phi &=& 4\pi G \rho \, . \label{PoissonEq}
\end{eqnarray}


\begin{deluxetable}{lcclcc}
\tabletypesize{\scriptsize}

\tablecaption{Initial Parameters for Model Q0.8\tablenotemark{\dag}} \tablewidth{0pt}

\tablehead{ \colhead{System} & \colhead{Initial} & ~ &
\colhead{Component} &
~ & ~ \\
\colhead{Parameter} & \colhead{SCF Value} & ~ & \colhead{Parameter}
& \colhead{Donor} & \colhead{Accretor}  }

\startdata
$q_0$ & 0.843 & ~~~~~~~~~~ & $M_i$ & 0.0223 & 0.0265   \\
$a_0$ & 0.9707 & ~~~~~~~~~~ & $\rho^\mathrm{max}_i$ & 0.8800 & 1.0000   \\
$\Omega_0$ & 0.2327 & ~~~~~~~~~~ & $K_i$ & 0.0387 & 0.0417   \\
$J_\mathrm{tot}$ & $2.93\times 10^{-3}$ & ~~~~~~~~~~ & $V_i$ & 0.1577 & 0.1634   \\
${R}_\mathrm{com}$ & $2.28\times 10^{-6}$ & ~~~~~~~~~~ & $V^\mathrm{RL}_i$ & 0.1788
& 0.2273   \\
\enddata
\tablenotetext{\dag}{Parameter values are given here in dimensionless
polytropic units. To scale these numbers to other (e.g., cgs) units, see the
discussion in Appendix~\ref{unitsAppendix}.}
\label{table1_0.843}
\end{deluxetable}

\section{Results of detached binary simulations}\label{sec_0.843_UB}
The two hydrocode modifications described in \S\ref{sec_corr_code}
were introduced to improve the accuracy and long-term stability of
our simulations so that we would be able to follow the dynamical
evolution of close binary systems through many orbits. In order to
assess the effects of these changes, we have carefully analyzed the
evolution of a binary system whose initial structure matched as
closely as possible the unequal-mass, detached binary system with
$q_0 = 0.843$ that was discussed in detail by MTF.  In MTF, this
``benchmark'' configuration was referred to as model ``UB'' (for
``unequal-mass binary''); we will henceforth refer to our closely
related initial configuration as model ``Q0.8,'' reflecting the
value of the model's initial mass-ratio.  For analysis purposes,
this same initial model, Q0.8, was evolved through approximately 5.3
orbits {\it four separate times} using slightly different versions
of the hydrodynamical code.  In what follows we first detail the
properties of this initial model, then we describe the results of
the four separate evolutions.

\begin{deluxetable}{crrrcc}
\tabletypesize{\scriptsize}

\tablecaption{Computational Grid Parameters} \tablewidth{0pt}

\tablehead{ \colhead{Model} & \colhead{$N_R$} & \colhead{$N_\phi$} &
\colhead{$N_z$} & \colhead{$R_\mathrm{grid}$} & \colhead{$\Delta
R$\tablenotemark{a}} }

\startdata
Q0.8 & 130 & 256 & 98 & 1.000 & $7.87\times 10^{-3}$   \\
Q1.3 & 162 & 256 & 98 & 1.251 & $7.87\times 10^{-3}$   \\
Q0.5 & 162 & 256 & 98 & 1.251 & $7.87\times 10^{-3}$
\enddata
\tablenotetext{a}{In all cases, $\Delta R = R_\mathrm{grid}/(N_R -
3)$; $\Delta z = \Delta R$; and $\Delta\phi = 2\pi/N_\phi$.}
\label{ComputationalGrid}
\end{deluxetable}

\begin{figure}[!ht]
\centering \includegraphics[scale=0.7]{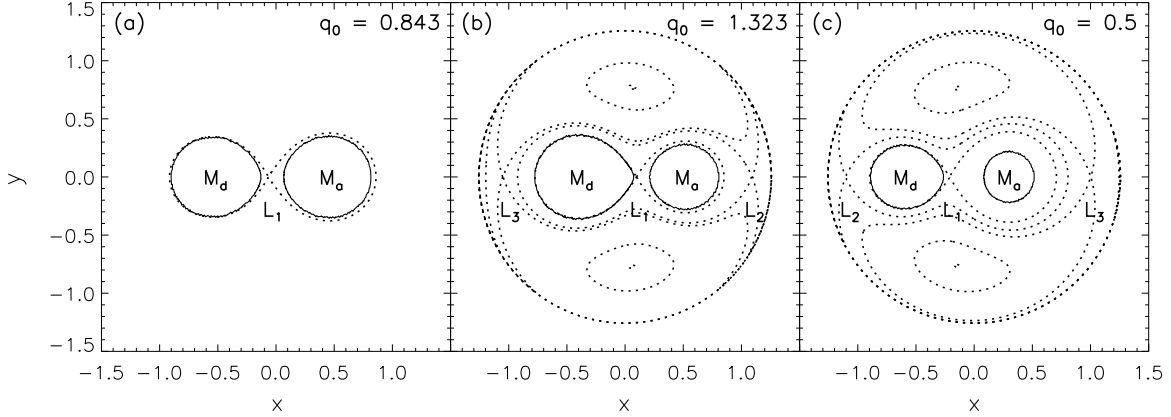} \caption{A slice through
the equatorial-plane is shown for (a) model Q0.8, (b) model Q1.3,
and (c) model Q0.5. In each panel, the solid, isodensity contours
denote the surfaces of the two stars; they are drawn at a level,
$\rho / \rho^{\mathrm{max}}_{a} = 10^{-5}$.  The ``donor'' star is
located on the left in each panel.  The dashed lines trace
selected equipotential contours of the self-consistently
calculated Roche potential for each model. The outermost circle
drawn in panels (b) and (c) identifies the radial edge,
$R_\mathrm{grid}$, of the computational grid that was used to
evolve these initial models hydrodynamically.  The location of the
inner ($L_1$) Lagrange point is identified for each system; for
models Q1.3 and Q0.5, two additional ($L_2$ and $L_3$) Lagrange
points are identified because they fall at positions $R <
R_\mathrm{grid}$.} \label{equipotentials}
\end{figure}

Figure \ref{equipotentials}$a$ shows the critical Roche surface
(dashed curve) and the equatorial-plane density contours (solid
curves) that trace the surface of the two stars in model Q0.8, as
constructed by our SCF code.  The inner (L1) Lagrange point on the
critical Roche surface is also identified.
Table \ref{table1_0.843} lists the numerical values of various
physical parameters that define this initial model. (Where
appropriate, the subscript ``0'' has been used to emphasize that
these are initial parameter values derived from the SCF code. All of
these physical variables were permitted to vary with time during the
course of the hydrodynamic evolutions.)  These include parameters
for the binary system as a whole
--- $q_0$, $a_0$, $\Omega_0$, $J_\mathrm{tot}$, and
$R_\mathrm{com}$ (as defined by Eq.~\ref{RcomDefined})
--- and parameters that define the structure of
the individual polytropic stars --- $M_i$, $K_i$,
$\rho^\mathrm{max}_i$, the volume occupied by each star $V_i$, and
the volume of the associated Roche lobe $V^\mathrm{RL}_i$, where the
subscript ``$i$'' refers either to the donor or the accretor.  We
note that, by volume, the less-massive star (destined to become the
donor) initially filled $88\%$ of its Roche lobe and the
more-massive star (the accretor) initially filled $72\%$ of its
Roche lobe.  We note as well that, initially, the center of mass of
this model was positioned almost exactly at the center of the
coordinate grid; specifically, relative to the initial binary
separation, $R_\mathrm{com}/a_0 = 2.35\times 10^{-6}$ initially.

All of the parameter values in Table \ref{table1_0.843} are given in
units such that $G = R_\mathrm{SCF} =\rho^\mathrm{max}_{\rm a}(t=0)
= 1$, where $R_\mathrm{SCF}$ is the outer edge of the cylindrical
grid that was used to generate the model in the SCF code.
Appendix~\ref{unitsAppendix} provides expressions that can be used to scale the values of
these dimensionless parameters to more meaningful ({\it e.g.}, cgs)
units.

As is detailed in the first row of Table \ref{ComputationalGrid},
model Q0.8 was evolved on a uniform grid that had a resolution of
(130, 256, 98) zones in $(R, \phi, z)$, respectively. The hydrocode
grid had the same radial extent, $R_\mathrm{grid}$, and the same
size grid zones, $\Delta z = \Delta R = R_\mathrm{grid}/(N_R-3)$ as
the grid that was used by the SCF code to construct the initial
model, that is, $R_\mathrm{grid} = R_\mathrm{SCF} = 1$ and $\Delta R
= 7.87\times 10^{-3}$.  As was mentioned above, we evolved model
Q0.8 through approximately 5.3 orbits, four separate times, using
slightly different versions of the hydrodynamical code. The versions
of the code differed, as follows (see \S\ref{sec_corr_code} for
additional clarification):
\begin{itemize}
\item Evolution Q0.8-H: The pressure source term in Euler's
equation was calculated using $\rho{\nabla} H$, and no correction
was made to limit the motion of the system's center-of-mass. (This
is identical to the manner in which source terms were handled in the
``benchmark'' UB model evolution published by MTF.)

\item Evolution Q0.8-P: The pressure source term in Euler's
equation was calculated using $\nabla p$, instead of $\rho{\nabla}
H$, but no correction was made to limit the motion of the system's
center-of-mass.

\item Evolution Q0.8-HC: A very small, artificial acceleration was
applied in an attempt to minimize the motion of the system's
center-of-mass, but the pressure source term in Euler's equation was
calculated using $\rho{\nabla} H$.

\item Evolution Q0.8-PC: The pressure source term in Euler's
equation was calculated using $\nabla p$, instead of $\rho{\nabla}
H$, and a very small, artificial acceleration was applied in an
attempt to minimize the motion of the system's center-of-mass.
\end{itemize}
Columns 2 and 3 of Table \ref{TableUB} highlight the differences
between these four model evolutions: a ``yes'' in column 2 means
that $\nabla p$ was used in place of $\rho\nabla H$; and a ``yes''
in column 3 means that ${\bf a}^\mathrm{art}$ was applied in an
effort to correct for the small, but undesired center-of-mass
motion.

Although both stars initially filled a large fraction of their
respective Roche volumes, neither star was actually in contact with
its Roche lobe initially.  Hence, the system was not initially
susceptible to a mass-transfer instability and nothing was done
during these four simulations to artificially excite the
instability. As has been explained by MTF, the evolution of this
``benchmark'' model serves to illustrate how well the fully
dynamical hydrocode can preserve the detailed structure of a
complex, equilibrium configuration that is close to, but has not
exceeded, contact.

\begin{figure}[!ht]
\centering \includegraphics[scale=0.8,viewport=12 16 540 650,clip]{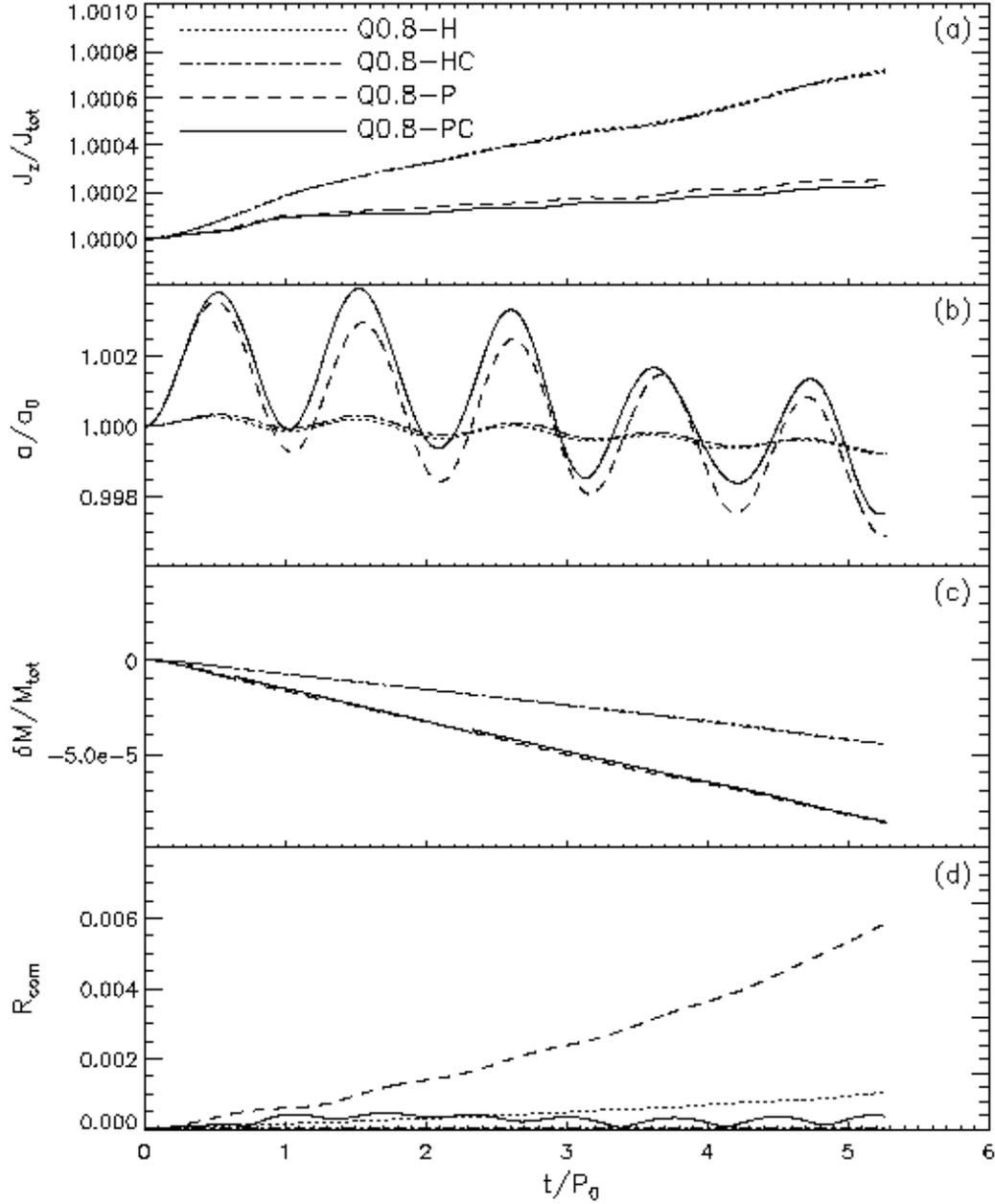} \caption{Results from the
evolution of model Q0.8 using four slightly different versions of
the hydrocode (evolution H -- dotted curve; HC -- dash-dot
curve; P -- dashed curve; PC -- solid curve). From top to
bottom: (a) The $z$-component of the system's angular momentum $J_z$,
normalized to its initial value $J_\mathrm{tot}$, is plotted as a
function of time $t$, normalized to the system's initial orbital
period $P_0 = 2\pi/\Omega_0$. (b) The binary separation $a$,
normalized to its initial value $a_0$, is plotted as a function of
$t/P_0$. (c) The difference between the total mass of the binary
system and its initial value, $\delta M \equiv M(t)-M_\mathrm{tot}$,
normalized to the initial mass, is plotted as a function of $t/P_0$.
(d) The distance in code units $R_{\rm com}$ from the axis of the
grid to the instantaneous center of mass as a function of $t/P_0$.
Note that the boundary of the innermost radial cell in these units
is at $R_{\rm com}=0.008$.} \label{Q0.8 plots}
\end{figure}

Column 4 of Table \ref{TableUB} records the length of time --- in
units of the initial orbital period, $P_0 = 2\pi/\Omega_0$ --- that
model Q0.8 was evolved in each of these benchmark simulations. In
accordance with our expectations, throughout all four model
evolutions, the binary system remained detached, the two stars moved
along very nearly circular orbits at a separation $a(t)$ that
deviated only slightly from the initial separation, and both stars
individually preserved their initial detailed, force-balanced
structures to a high degree of precision. The top three panels of
Figure \ref{Q0.8 plots} show how well key system parameters were
preserved throughout all four evolutions:  (a) the $z$-component of
the total angular momentum $J_{z}(t)$, measured relative to the
initial value, $J_\mathrm{tot}$; (b) the binary separation $a(t)$,
normalized to the initial separation $a_0$; and (c) the change in
total system mass $\delta M(t) \equiv \{M(t) - M_\mathrm{tot}\}$,
measured relative to the initial total mass $M_\mathrm{tot}$, where
$M(t)$ is the total system mass at time $t$. The separation $a(t)$
reported in all the simulations in this paper is the distance
between the centers of mass of the two components.  In detail, we
identify material belonging to each star by dividing the
computational grid into two regions 
with a plane perpendicular to the binary axis going through the
approximate location of the inner Lagrange point. Individual grid
cells contribute to the center of mass summation if they are more
strongly bound to their respective star than the surface layer (a
constant density surface). Finally, panel (d) shows the small drift
of the center of mass away from the center of the grid as a function
of time for the various runs. Note that in all cases the center of
mass remained within the first radial cell ($\Delta R=7.87\times
10^{-3}$) and that the drift is significantly reduced by the center
of mass corrections for both Q0.8-PC and Q0.8-HC, being almost
invisible for the latter.

As the curves in Figure \ref{Q0.8 plots}
show, over the course of approximately 5 orbits in all four model
Q0.8 simulations, the binary lost a very small fraction of its mass,
it slowly gained a very small amount of angular momentum and, at the
same time, it experienced a slow, very slight secular decay of the
orbit. In addition to the slight orbital decay, a small-amplitude
oscillation occurs in the binary separation with a period
approximately equal to the initial orbital period of the binary.
This ``epicyclic'' motion probably simply reflects that the orbit is
not precisely circular, in which case the mean amplitude of the
epicyclic oscillation, $(\Delta a/a)_\mathrm{epicyclic}$, can be
interpreted as the eccentricity of the orbit. The epicyclic motion
probably arises because, when the initial model was inserted into
the hydrodynamical code, its assigned angular velocity, $\Omega_0$,
was slightly different from the value that would have been required
to place the stars into a perfectly circular orbit. Columns 5 - 8 of
Table \ref{TableUB} list, respectively, the secular rate of change
per orbit of $J_{z}$, $a$ and the total system mass $M$, as well as
the epicyclic amplitude $(\Delta a/a)_\mathrm{epicyclic}$ that
resulted from each of our four model Q0.8 simulations; also listed
in the first row of the table are the corresponding values from
MTF's UB simulation.\footnote{We report here a typographical error
in the values quoted by MTF for $\Delta M_{1}/M_{\rm tot}$ and
$\Delta M_{2}/M_{\rm tot}$. The correct values are $\Delta
M_{1}/M_{\rm tot} \approx -3.0 \times 10^{-6}$ and $\Delta
M_{2}/M_{\rm tot} \approx -1.1 \times 10^{-5}$}

\begin{deluxetable}{lccccccc}
\tabletypesize{\scriptsize}

\tablecaption{Results from Detached Binary Evolutions (Model Q0.8)}
\tablewidth{0pt}

\tablehead{ \colhead{Simulation} & \colhead{$\nabla P$} &
\colhead{COM} & \colhead{$t/P_0$} & \colhead{$\Delta
J_z/J_\mathrm{tot}$} & \colhead{$(\Delta a/a)_\mathrm{secular}$} &
\colhead{$(\Delta M/M_\mathrm{tot})$} & \colhead{$(\Delta
a/a)_\mathrm{epicyclic}$} \\

\colhead{(1)} & \colhead{(2)} & \colhead{(3)} & \colhead{(4)} &
\colhead{(5)} & \colhead{(6)} & \colhead{(7)} & \colhead{(8)} }

\startdata
UB      & no  & no  & 5.178 & $+1.5\times 10^{-4}$ & $-1.9\times 10^{-4}$  &
$-1.4\times 10^{-5}$ & $2.2\times 10^{-4}$  \\
Q0.8-H  & no  & no  & 5.263 & $+1.4\times 10^{-4}$ & $-1.5\times 10^{-4}$  &
$-8.6\times 10^{-6}$ & $2.0\times 10^{-4}$  \\
Q0.8-HC & no  & yes & 5.265 & $+1.4\times 10^{-4}$ & $-1.5\times 10^{-4}$  &
$-8.5\times 10^{-6}$ & $2.0\times 10^{-4}$  \\
Q0.8-P  & yes & no  & 5.269 & $+4.7\times 10^{-5}$ & $-6.1\times 10^{-4}$  &
$-1.6\times 10^{-5}$ & $2.2\times 10^{-3}$  \\
Q0.8-PC & yes & yes & 5.261 & $+4.4\times 10^{-5}$ & $-4.6\times 10^{-4}$  &
$-1.6\times 10^{-5}$ & $2.0\times 10^{-3}$  \\
\enddata
\label{TableUB}
\end{deluxetable}

In comparing the detailed results of the four Q0.8 evolutions, as
summarized in Table \ref{TableUB}, we notice that simulations that
use the gradient of the pressure conserve angular momentum slightly
better (to a level of 5 parts in $10^5$, per orbit, instead of 14
parts in $10^5$) than simulations in which the gradient of the
enthalpy was used. At the same time, however, the amplitude of the
epicyclic motion was roughly an order of magnitude larger (the
orbital eccentricity was $\approx 2\times 10^{-3}$ instead of
$\approx 2\times 10^{-4}$) and the rate of mass loss was
approximately twice as large (mass was conserved to a level of $16$
parts in $10^6$, per orbit, instead of $8$ parts in $10^6$) in the
simulations that used the gradient of the pressure. As was noted in
\S\ref{sec_corr_code}, expressing the pressure source term in
Euler's equation in terms of the gradient of the pressure provides
the more physically correct description of the evolution of
unequal-mass binaries undergoing mass transfer.  This may be the
reason evolutions Q0.8-P and Q0.8-PC conserved angular momentum
better than simulations that used the gradient of the enthalpy. In
the context of the difference in the epicyclic amplitude for the two
implementations we recall that the SCF code uses the gradient of the
enthalpy, rather than the pressure, to construct the initial state.
When this initial configuration is placed into the version of the
hydrocode that implements the $\rho\nabla H$ source term (evolutions
Q0.8-H and Q0.8-HC), no resultant forces are introduced and the
equilibrium of the initial model is therefore well preserved.
However, this will not be the case for the version of the hydrocode
that evolves the fluid with $\nabla p$. Although, strictly speaking,
$\nabla p = \rho\nabla H$ in an isentropic fluid (such as the fluid
in either one of the stars in this detached binary), in practice the
finite-difference expressions for these two gradients produce
slightly different values of the source term.  We suspect that the
$\nabla p$ implementation introduces very small, but nonzero
accelerations throughout the interior of both stars that cause the
two stars to oscillate slightly about their own, ideal equilibrium
configurations, and that it is the coupling between these stellar
oscillations and orbital dynamics of the system as a whole that
excites the larger epicyclic motions in evolutions Q0.8-P and
Q0.8-PC.

The effect of the center of mass correction was less dramatic than
implementing the gradient of the pressure or enthalpy; it slightly
improved the conservation of angular momentum and also slightly
reduced the amplitude of the epicyclic motion. However, the total
mass of the binary, on the scale presented here, remained virtually
unchanged for simulations with or without the center of mass
correction.

The three principal spurious effects --- the slow decay of the
orbit, the slow gain of angular momentum and the slow loss of mass
from the system --- seen in MTF's UB simulation are still present in
the simulations presented here. However, depending on the version of
hydrodynamical code we choose, our simulations show that the size of
these effects can be modified somewhat. The criterion for choosing
one version over another rests on the type of binary system we want
to simulate. For detached/semi-detached binaries having similar
components with respect to the polytropic constant, $\nabla H$ will
correctly describe the evolution even if mass transfer occurs in the
system. On the other hand if the components have different
polytropic constants ({\it i.e.}, the fluid in the two stars has
different specific entropies) then the $\nabla p$ scheme needs to be
used, particularly for binary systems that undergo mass-transfer.
The disadvantage of using the $\nabla p$ scheme is that a relatively
large epicyclic motion is induced in the orbit, and mass is
conserved to a slightly lower degree of accuracy than when the
$\nabla H$ scheme is utilized. The former can be reduced by
modifying the angular velocity to place the initial model in a
nearly circular orbit or by ``tweaking'' the initial state so as to
balance the forces which arise from implementing the $\nabla p$
scheme. In all five of the mass-transfer simulations that we present
below in \S\ref{sec_mdot_results}, the ``PC'' version of the
hydrocode has been used (consistent with the set of equations that
has been summarized above in \S\ref{equations}) because this version
is best suited to follow evolutions through many orbits when the
material residing in the donor has a specific entropy that differs
from the material initially making up the accretor.

The number recorded in column 7 of the last row of Table
\ref{TableUB} provides a reasonable estimate of the lowest
mass-transfer rate that we will be able to resolve with our existing
hydrocode. Specifically, we should expect to only be able to resolve
mass-transfer rates $\dot{M}_d(P_0/M_\mathrm{tot})
> \dot{M}_\mathrm{min} = - 1.6\times 10^{-5}$ because, as has just
been demonstrated, even a detached binary system like our model Q0.8
slowly loses mass at this rate. In order to accommodate this
restriction in our simulations of mass-transferring binary systems,
we will be required to ``drive'' the donor into sufficiently deep
contact with its Roche lobe that mass transfer proceeds at a rate
that exceeds $\dot{M}_\mathrm{min}$.  For MS Algol-type binaries
($M_\mathrm{tot}\approx 5 M_\odot, P_0 \approx 3~\mathrm{days}$)
\citep{BRM95}, this limit is equivalent to $\sim 10^{-2}
M_\odot/\mathrm{yr}$; and for AM CVn-type DWD binaries
($M_\mathrm{tot}\approx 1 M_\odot, P_0 \approx 0.3~\mathrm{hr}$)
\citep{PHS93, NPVY01}, this limit is equivalent to $\sim 0.4
M_\odot/\mathrm{yr}$.  Both of these limits are orders of magnitude
larger than the stable mass-transfer rates that are observed in
Algol-type or AM CVn-type binary systems. Hence, as we remarked in
the introductory section of this paper, we are unable to model
long-term, stable evolutionary phases of mass-transfer in such
systems.  However, during phases of unstable mass-transfer
--- which are the focus of our present study ---
the rate of mass-transfer can easily climb to levels above
$\dot{M}_\mathrm{min}$ and can be satisfactorily modelled with our
hydrocode. Note that the minimum resolvable mass transfer
$\dot{M}_\mathrm{min}$ is on the order of the Eddington critical
rate for both white dwarfs and main-sequence donors, but to treat
radiative forces is beyond the scope of this paper. Analytic
considerations \citep{HaWe99} suggest that the effect of mass loss
occurring at super-Eddington mass-transfer rates is stabilizing,
thus ignoring the radiative forces at this stage provides a more
stringent test of stability. Ideally one would like to be able to
follow the evolution with all relevant physical effects included and
to resolve realistic mass transfer rates, but this is not possible with
the computational resources available today.

\section{Mass-Transfer Simulations}\label{sec_mdot_results}
We now present results from mass-transfer simulations of two
semi-detached, $n=3/2$ polytropic binary systems with initial mass
ratios, $q_0 = 1.323$ and $q_0 =0.5$; these will henceforth be
referred to as models Q1.3 and Q0.5, respectively.  Paralleling the
information provided in Table \ref{table1_0.843} for model Q0.8,
Tables \ref{table1_1.3} and \ref{table0.5} list system parameters as
well as information about the structure of the component stars for
these two initial models, and Figures \ref{equipotentials}b and
\ref{equipotentials}c illustrate their initial equatorial-plane
structures. In both panels, the donor is the star on the left which
nearly fills its critical Roche surface. The outermost circle in
Figures \ref{equipotentials}b and \ref{equipotentials}c identifies
the edge of the computational grid on which both of these models
were evolved.  In contrast to our simulations of model Q0.8, and as
is summarized in Table \ref{ComputationalGrid}, the edge of the
hydrocode grid ($N_R = 162; R_\mathrm{grid} = 1.251$) was extended
beyond the grid that was used in the SCF code ($N_R = 130;
R_\mathrm{SCF} = 1.000$) in order to ensure that the L2 and L3
Lagrange points both were included in the computational domain.
Material that flows radially outward across either one of these two
saddle points in the effective potential is unlikely to return to
either one of the stars on a dynamical timescale, but the dynamical
motions of material that lies inside the L2 and L3 locations should
be fully included in the hydrodynamical simulations.

These particular models were chosen for this investigation, in part,
because their values of $q_0$ fall into two separate stability
regimes, as outlined above in \S \ref{SecExpectations}. For model
Q1.3, $q_0 > 1$ so the binary is expected to be violently unstable
to mass transfer.  For model Q0.5, $q_0$ falls below the value
$q_\mathrm{stable} = 2/3$ at which binaries with $n=3/2$ polytropic
structures should become stable against mass transfer
\citep{HaWe99}, if the effect of direct-impact accretion is ignored.
In addition, the properties of the stars in these two models were
specified in such a way that they permit us to examine two
distinctly different, but astrophysically interesting evolutionary
scenarios: In model Q1.3 (see Table \ref{table1_1.3}), the ratio of
the effective stellar radii $R_d/R_a = (V_d/V_a)^{1/3} \approx
M_d/M_a$, which mimics the mass-radius relationship of MS stars; and
in model Q0.5 (see Table \ref{table0.5}), $K_d = K_a$ and
$(R_d/R_a)^3 = (V_d/V_a) \approx M_a/M_d$, which represents well the
structural properties and mass-radius relationship of stars in a
low-mass DWD system. Finally, in the $q <2/3$ parameter regime, we
specifically selected an initial model with $q_0 = 0.5$ in order to
permit a direct comparison with one of the SPH simulations that was
reported by RS95.

In both models, as they were generated by the SCF code, the donor
star slightly underfills its Roche lobe initially.  Specifically,
for model Q1.3 (see Table \ref{table1_1.3}), $R_d/R^\mathrm{RL}_d =
(V_{\mathrm{d}}/V^{\mathrm{\rm RL}}_{\mathrm{d}})^{1/3} = 0.989$,
and for model Q0.5 (see Table \ref{table0.5}), $R_d/R^\mathrm{RL}_d
= (V_{\mathrm{d}}/V^{\mathrm{\rm RL}}_{\mathrm{d}})^{1/3} = 0.965$.
This means that neither model was actually undergoing mass-transfer
when it was introduced into the hydrocode. In the context of the SCF
technique, it is possible to iteratively refine the converged model
configuration to produce semi-detached binaries that come
progressively closer to filling their self-consistently determined
Roche volume, up to some limiting volume-filling factor $\lesssim
1.00$ that is set by the finite grid resolution of the SCF code. But
the SCF technique will not converge to a model where stellar
material extends beyond the Roche lobe as this material can not be
in hydrostatic equilibrium. Even if the SCF code were able to
generate models where the donor star completely fills its critical
Roche surface, the models would have little additional practical
utility over models, such as Q1.3 and Q0.5, where $R_d$ is $97 -
99\%$ of $R_d^\mathrm{RL}$. This is because such models would only
be marginally unstable to mass transfer so the time that would be
required for $\dot{M}_d$ to grow to levels that are of interest (or
even resolvable) in the present investigation would be prohibitively
long for a time-explicit hydrodynamics scheme such as ours.  It is
therefore necessary for us to employ some means of artificially
``driving'' each binary model into deep enough contact with its
Roche surface to ensure that, early in the hydrodynamical evolution,
the mass transfer rate climbs to a rate that is resolvable, that is,
to a rate $\gtrsim \dot{M}_\mathrm{min}$.  After this has been
achieved, the driving is turned off, except in one case of the Q0.5
runs in which the driving was applied throughout the entire
eveolution.

One can imagine several possible mechanisms for driving a binary
system into contact.  For example, RS95 began with two widely
separated, spherical stars and slowly damped the orbital velocity
with an artificial friction term until the donor began to lose
particles through the inner Lagrange point. This technique is
particularly well adapted to the SPH method, as it is free of a
computational grid.  We instead begin with tidally deformed,
rotationally flattened equilibrium models that are already very near
contact. To proceed from this point we could simply impart an inward
radial kick to the donor, but this would certainly cause the orbit
to immediately become noncircular.  Alternatively, we could attempt
to mimic natural processes by (a) forcing the donor to slowly expand
to fill its Roche lobe --- a process that will occur in MS binaries
as the more massive star evolves off the main sequence --- or (b)
slowly removing orbital angular momentum, thereby forcing the Roche
lobe to contract around the donor
--- as occurs in DWD systems as angular momentum is lost
from the system via gravitational radiation.  Ideally, the outcome
of a given mass-transfer instability should be insensitive to the
evolutionary process that has brought the donor star into contact
with its critical Roche surface.

In our present investigation, we have explored both of these more
natural processes as a means of driving the donor into sufficiently
deep contact with its critical Roche surface so that $\dot{M}_d$
climbs above $\dot{M}_\mathrm{min}$.  Of course, because the
hydrodynamical evolutions have been followed with an explicit
time-integration scheme, we have found it necessary to ``drive'' the
donor into contact at a rate that far exceeds the natural thermal
expansion rate of evolving MS stars or the natural rate at which
angular momentum is lost from DWD binaries.  The rates we have
employed are, nevertheless, slow enough that during the early (and
generally brief) phase of artificial driving, the two stars
individually, as well as the system as a whole, remain very near the
initial equilibrium state as generated by the SCF code.  Thus, as
desired, the artificial driving introduces only secular, rather than
dynamical, changes in the system.

In order to initiate mass transfer in model Q0.5 (see \S
\ref{sec_0.5_results}), we drained orbital angular momentum from the
system at a rate of 1\% of $J_\mathrm{tot}$ per orbit for varying
amounts of time in order to achieve varying depths of contact and
levels of mass transfer. These simulations will henceforth be
collectively referred to as Q0.5-D and individually as Q0.5-Da, -Db,
and -Dc in order of increasing driving time (see
\S\ref{sec_0.5_results} for details). For model Q1.3 (see \S
\ref{sec_1.3_results}), we tried both mechanisms: In simulation
Q1.3-D, we drained orbital angular momentum from the system at a
rate of 1\% of $J_\mathrm{tot}$ per orbit for two orbits; in
simulation Q1.3-E, we forced a slow expansion of the donor by
increasing $K_d$ at a rate of 1.67\% of its initial value over a
timescale of 1 orbit for the first two orbits
(see Appendix~\ref{drivingAppendix} for
details). A detailed comparison of simulations Q1.3-D and Q1.3-E
confirms that the outcome of the instability is insensitive to the
process by which the donor has been brought into contact with its
critical Roche surface.

\subsection{Simulations with $M_{\mathrm{d}} = 1.323 M_{\mathrm{a}}$ Initially}
\label{sec_1.3_results}

\subsubsection{Evolution Q1.3-D}

\begin{deluxetable}{lcclcc}
\tabletypesize{\scriptsize}

\tablecaption{Initial Parameters for Model Q1.3\tablenotemark{\dag}}
\tablewidth{0pt}

\tablehead{ \colhead{System} & \colhead{Initial} & ~ &
\colhead{Component} &
~ & ~ \\
\colhead{Parameter} & \colhead{SCF Value} & ~ & \colhead{Parameter}
& \colhead{Donor} & \colhead{Accretor}  }

\startdata
$q_0$ & 1.323 & ~~~~~~~~~~ & $M_i$ & 0.0176 & 0.0133   \\
$a_0$ & 0.8882 & ~~~~~~~~~~ & $\rho_i^\mathrm{max}$ & 0.6000 & 1.0000   \\
$\Omega_0$ & 0.2113 & ~~~~~~~~~~ & $K_i$ & 0.0372 & 0.0264   \\
$J_\mathrm{tot}$ & $1.40\times 10^{-3}$ & ~~~~~~~~~~ & $V_i$ & 0.1810 & 0.0799   \\
${R}_\mathrm{com}$ & $7.11\times 10^{-5}$ & ~~~~~~~~~~ & $V_i^\mathrm{RL}$ & 0.1869
& 0.1261   \\
\enddata
\tablenotetext{\dag}{Parameter values are given here in dimensionless
polytropic units. To scale these numbers to other (e.g., cgs) units, see the
discussion in Appendix~\ref{unitsAppendix}.}
\label{table1_1.3}
\end{deluxetable}

\begin{figure}[!ht]
\centering \includegraphics[scale=0.8,viewport=67 498 521 719,clip]{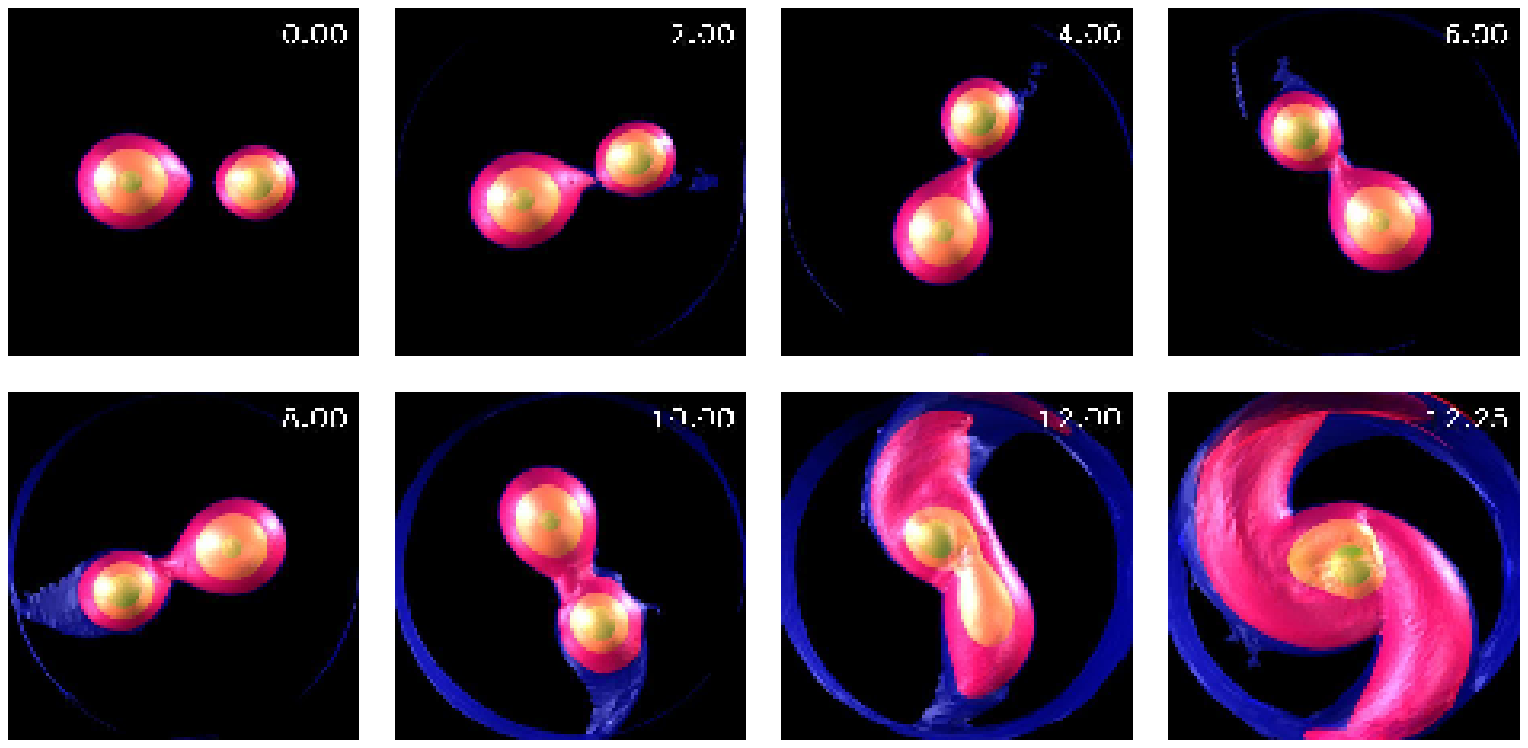} \caption{A
three-dimensional rendering of the mass density at selected time
slices during the Q1.3-D evolution that demonstrate direct impact
accretion. Times are shown on the top right of each frame in units
of the initial orbital period. Each image is a snapshot as viewed by
an observer looking down on the equatorial plane while rotating with
the angular velocity $\Omega_0$. The four colored transparent shells
are iso-density surfaces shown at levels $\rho/ \rho^{\mathrm{\rm
max}}_{\rm a}$ = 0.5 (green), 0.1 (yellow), $10^{-3}$ (red) and
$10^{-5}$ (blue) where $\rho^{\mathrm{\rm max}}_{\rm a}$ is the
maximum density of the accretor --- the star on the right in the
initial frame (t = 0) of the figure. In the online version this
figure is supplemented by a mpeg animation showing the full
evolution.} \label{Q1.3-D images}
\end{figure}

Figure \ref{Q1.3-D images} presents a sequence of images showing the
structure of the binary at selected times during the Q1.3-D
evolution. Each image is a three-dimensional rendering of the
mass-density distribution as viewed looking down on the equatorial
plane of the binary from a frame of reference rotating with the
initial orbital frequency, $\Omega_0$. The four nested iso-density
surfaces have been drawn at levels $\rho/ \rho^{\mathrm{\rm
max}}_{a} = 0.5$ (green), 0.1 (yellow), $10^{-3}$ (red) and
$10^{-5}$ (blue), where $\rho^{\mathrm{\rm max}}_{a}$ is given in
Table \ref{table1_1.3}. The eight images presented in Figure
\ref{Q1.3-D images} have been taken from an animation sequence that
includes approximately 1500 frames (120 frames per orbital period)
and illustrates in much more detail the dominant structures that
developed during the Q1.3-D evolution.  The first seven images
displayed here are equally spaced in time at intervals of $2P_0$.
The gradual, counter-clockwise
shift in the position angle of the line that connects the centers of
the two stars reflects the gradual, monotonic increase in the
system's orbital angular velocity $\Omega$, which in turn reflects a
gradual decrease in the system's orbital separation, $a$.

\begin{figure}[!ht]
\centering \includegraphics[scale=0.8,viewport=12 16 465 592,clip]{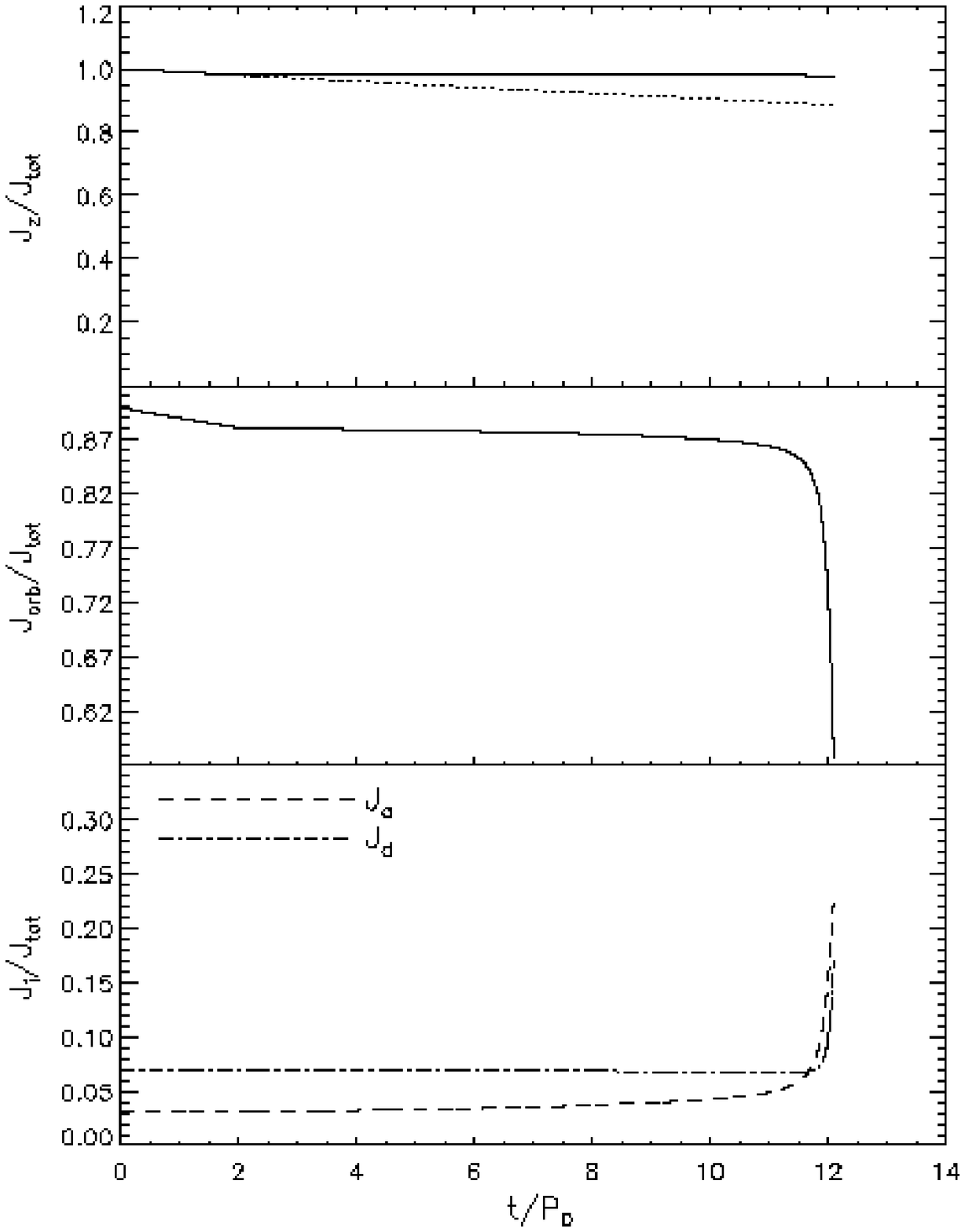} \caption{Top: The
$z$-component of the total angular momentum with respect to an
inertial frame centered at the center of mass, $J_{z}$, normalized
to its initial value for the Q1.3-D (solid curve).  The initial loss
rate of angular momentum at 1\% per orbit is denoted by the dotted
line.  The driving for Q1.3-D is stopped after two orbits and this
can be seen when the solid curve becomes horizontal. Middle and
bottom: evolution of the orbital angular momentum and the spin
angular momenta of the components all normalized to the total
initial angular momentum for the Q1.3-D run.} \label{Q1.3-D J plots}
\end{figure}

By the end of the first two orbits, an accretion stream has begun to
develop. We quickly appreciate that, for the selected mass-radius
relationship of this binary system, the accretor has a sufficiently
large radius that the mass-transfer stream must directly impact,
rather than go into orbit around, the accretor. The material is
moving supersonically as it hits the accretor's surface. This
creates a standing shock wave of relatively high thermal-pressure
material on the accretor's surface (visible as a ridge of material
in the images shown at $t/P_0 =$ 6, 8, and 10) that balances the ram
pressure of the in-falling donor material. The stream clearly
strikes the surface of the accretor at an oblique angle.  Hence,
while the stream's motion perpendicular to the surface is abruptly
halted by the standing shock front, its motion tangent to the
surface continues unabated.  In this manner, angular momentum is
added to the spin of the accretor.  It is precisely this complex,
three-dimensional flow --- along with the tidal torques generated by
the perturbed density distribution --- that characterizes
direct-impact accretion and necessitates a fully three-dimensional
hydrodynamic treatment to quantify its role.

Throughout the Q1.3-D evolution, the accretion stream steadily grows
thicker, reflecting a steady increase in the rate at which material
from the donor is being transferred to the accretor.  By the time
the system has completed twelve orbits, it can no longer be
described as two stars in nearly circular orbits that are loosely
tied together by a mass-transfer stream.  Instead, the two stars
have begun to plunge toward one another and, only one quarter of an
orbit later (see Figure \ref{Q1.3-D images}) the central cores of
the two stars have merged.  This final, short merger phase of the
evolution results from what \citet{LRS3} have identified as a tidal
instability.  At late times a fairly steady stream of very low
density material (identified by the blue iso-density surface) also
emerges from the trailing edge of the less massive star, that is,
the accretor, flows out through the $L_2$ Lagrange point, and
accumulates along the boundary of the computational grid. In this
manner, $ \sim 0.1\%$ of the total system mass is lost from the
system during the first 12 orbits --- a remarkably high level of
conservation through the mass transfer event, given the very high
mass-transfer rate attained prior to merger.

\begin{figure}[!ht]
\centering \includegraphics[scale=0.8,viewport=12 16 465 592,clip]{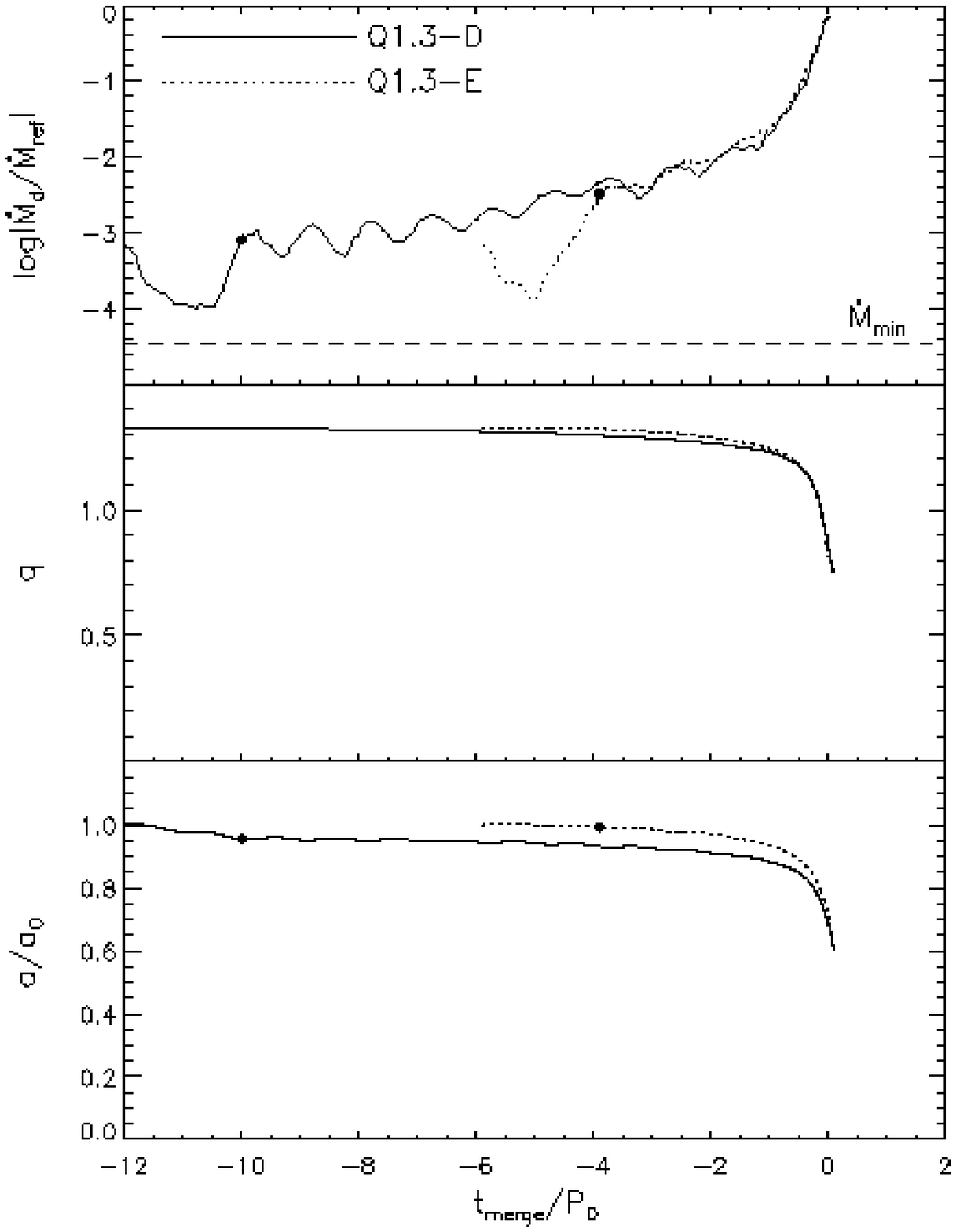} \caption{The evolution of
the mass-transfer rate (top), the mass-ratio $q$, and the separation
$a$ for the same initial binary driven to contact by removing
angular momentum (Q1.3-D, solid curves) and by expansion of the
donor (Q1.3-E, dotted curves). The times shown are times to `merger'
measured in initial binary periods. The dashed line in the top panel
indicates the minimum level of mass transfer ${\dot M}_{\rm min}$
that is resolvable in our simulations (see text). The separation $a$
is normalized to its initial value $a_0$. The mass transfer rate is
normalized to the reference value $\dot M_{\rm ref}\equiv M_{\rm d}
(0)/P_0$ (initial donor mass divided by initial orbital period). For
most of the evolution $q>1$, so $a$ decreases as a result of mass
transfer even in the absence of driving. The dots mark the points at
which driving was stopped.} \label{Q1.3-D-E Mqa plots}
\end{figure}

For purposes of further discussion it is productive to divide the
Q1.3-D evolution into three phases that are contiguous in time: the
``driving'' phase ($0\leq t/P_0 \leq 2$); the ``mass-transfer''
phase ($2\leq t/P_0 \leq 12$); and the final ``merger'' phase  ($12
\leq t/P_0 \leq 12.25$).  More quantitative descriptions of the
first two of these evolutionary phases are provided by Figures
\ref{Q1.3-D J plots} and \ref{Q1.3-D-E Mqa plots}. Figure
\ref{Q1.3-D J plots} shows the time-dependent behavior of (top
panel, solid line) the $z$-component of the system's total angular
momentum, (middle panel) the system's orbital angular momentum, and
(bottom panel) the spin angular momentum of both the donor
(dot-dashed curve) and the accretor (dashed curve). In all three
panels of Figure \ref{Q1.3-D J plots}, the quantity being plotted is
normalized to the system's initial total angular momentum,
$J_\mathrm{tot}$, and the curves have not been extended beyond
$t/P_0 = 12$ because at later times it becomes difficult to
distinguish between the individual stellar components. The solid
curves in Figure \ref{Q1.3-D-E Mqa plots} display the time-dependent
behavior of the mass transfer rate $\dot{M}_d$ normalized to the
ratio $\dot M_{\rm ref}\equiv M_d(t=0)/P_0$ (top panel), the system
mass ratio $q$ (middle panel), and the orbital separation $a$
normalized to $a_0$ (bottom panel). For reasons that will become
apparent, below, in Figure \ref{Q1.3-D-E Mqa plots} the time
coordinate (horizontal axis) has been shifted to
$t_\mathrm{merge}\equiv (t_{\rm D} - 12P_0)$, where $t_{\rm D}$ is
the evolutionary time recorded during the Q1.3-D simulation, so that
the onset of the final merger phase occurs at the origin ({\it
i.e.}, at $t_\mathrm{merge} = 0$).

As was explained earlier, during the ``driving'' phase of the Q1.3-D
evolution, angular momentum was artificially extracted from the
system at a rate of 1\% of $J_\mathrm{tot}$ per orbit.  This is
directly reflected in the top panel of Figure \ref{Q1.3-D J plots},
where the behavior of $J_z(t)$ can be compared with a (dotted) line
whose slope is precisely $0.01 J_\mathrm{tot}/P_0$. During this
phase of the evolution, angular momentum was extracted from the
system in such a way that the spin angular momenta of the two
stellar components $J_d$ and $J_a$ (bottom panel of Figure
\ref{Q1.3-D J plots}) remained essentially unchanged.
(See Appendix~\ref{drivingAppendix}
for a precise description of this extraction method.) In practice,
then, the extraction of angular momentum resulted in a steady drop
in the system's orbital angular momentum over the first two orbits
(see the middle panel of Figure \ref{Q1.3-D J plots}) as, in a
terminology that is consistent with the discussion associated with
Eq.~(\ref{adot}), the system experienced a systemic loss of angular
momentum $(\dot{J}/J_\mathrm{orb})_\mathrm{sys} \approx 0.01/P_0$.
According to Eq.~(\ref{adot}), therefore, it is no surprise that
during the driving phase of this evolution the orbital separation
decreased at a rate of $\dot{a}/a_0 \approx 0.02/P_0$ (see the
bottom panel of Figure \ref{Q1.3-D-E Mqa plots}), that is, to a
value $a/a_0 \approx 0.96$ after two orbits.  This, in turn,
produced an $\approx 4\%$ reduction in the effective Roche-lobe
radii, which produced the desired result of bringing the donor's
Roche lobe into a sufficiently deep contact with the stellar surface
of the donor that the mass-transfer event was initiated. As is shown
in the top panel of Figure \ref{Q1.3-D-E Mqa plots}, at the end of
this ``driving'' phase of the evolution (marked by the solid dot),
$|\dot{M}_d| \approx 10^{-3} M_d/P_0$. This mass-transfer rate was
sufficiently large (compared to $\dot{M}_\mathrm{min}$, for example)
that it produced a recognizable  and spatially resolvable accretion
stream (Figure \ref{Q1.3-D images}).

At the end of the driving phase of the evolution, we stopped
extracting angular momentum from the system; in effect, the external
driving that had been producing a systemic loss of angular momentum
$(\dot{J}/ J_\mathrm{orb})_\mathrm{sys}$ was set to zero. Throughout
the remainder of the Q1.3-D evolution, therefore, the system's total
angular momentum was conserved to a high level of precision (see the
top panel of Figure \ref{Q1.3-D J plots}, where the solid curve is
perfectly horizontal), and variations in the orbital separation
could only be attributed to the internal {\it redistribution} of
angular momentum associated with mass transfer and tides, as is
approximately described by the last two terms on the right-hand-side
of Eq.~(\ref{adot}).

A few key features are identifiable in the plot of $a(t)$ during the
``mass-transfer'' phase of the Q1.3-D evolution (solid curve in the
bottom panel of Figure \ref{Q1.3-D-E Mqa plots}).  First, the
separation slowly, but steadily decreases in step with the slow,
steady decrease of the system mass ratio $q$ (solid curve in the
middle panel of Figure \ref{Q1.3-D-E Mqa plots}).  This appears to
be in accord with the explicit $q$-dependence of the last term on
the right-hand-side of Eq.~(\ref{adot}).  Second, as was observed in
the benchmark evolutions of model Q0.8, $a(t)$ displays a
low-amplitude oscillation with a period $\approx P_0$.  This
``epicyclic'' oscillation reflects the fact that the orbit is
slightly noncircular.  Although there is evidence that this
epicyclic motion was amplified somewhat during the ``driving'' phase
of this evolution, it is perhaps significant that the eccentricity
of the orbit did not noticeably grow during the mass-transfer phase.
Third, in association with the onset of a tidal instability, $a(t)$
begins to decrease rapidly as $t_\mathrm{merge}$ approaches zero.

As the solid curve in the top panel of Figure \ref{Q1.3-D-E Mqa
plots} shows, during the driving phase of the Q1.3-D evolution the
accretion rate remains low as the surface of the donor is being
brought closer to, and then into deeper contact with its Roche lobe.
However, after the mass-transfer phase begins in ernest, the
accretion rate steadily increases. This is as expected because,
although the mass ratio $q$ steadily decreases throughout the
mass-transfer phase of the evolution, it remains larger than unity
until the rapid plunge and merger phase gets underway. (As shown in
the middle panel of Figure \ref{Q1.3-D-E Mqa plots}, $q > 1$ until
$t/P_0 = 11.884$.) Hence, throughout most of this evolution,
$R_d^\mathrm{RL}$ steadily decreases while $R_d$ steadily increases,
so a larger and larger fraction of the donor's envelope rises above
the critical Roche surface.  This, in itself, is a formula for
disaster and is sufficient to explain why merger was the inevitable
outcome of this $q_0 > 1$ mass-transfer evolution. It should be
noted as well that, as discussed in \S \ref{SecDirectImpact} above,
the phenomenon of direct-impact accretion (Marsh \& Steeghs 2002;
Marsh et al. 2004) must also have acted to further destabilize mass
transfer in this system.

Throughout the mass-transfer phase of this evolution, $\dot{M}_d$
exhibits a low-amplitude oscillation on top of its steady, secular
rise. This oscillation has a period $\approx P_0$, strongly
suggesting that it is associated with the epicyclic oscillation seen
in the plot of $a(t)$.  This association is substantiated by the
realization that, during each oscillation when $\dot{M}_d$ reaches a
local maximum, the orbital separation $a$ is at a local minimum.
That is, the simulation permits us to see variations in the
mass-transfer rate that are directly associated with the slightly
non-circular shape of the orbit.  Because the eccentricity of the
orbit ({\it i.e.}, the amplitude of the epicyclic oscillations) does
not noticeably increase during the evolution, whereas the
mass-transfer rate does steadily increase, it is understandable that
the relative amplitude of the oscillations that are seen in the plot
of $\dot{M}_d(t)$ decrease with time. During the final two orbits of
this evolution, the mass-transfer rate climbs rapidly to a very high
amplitude, reaching a level $\gtrsim 10\% (M_d/P_0)$ shortly before
the final merger phase. This behavior of $\dot{M}_d(t)$ during the
last portion of the mass-transfer phase of evolution Q1.3-D strongly
resembles the behavior that has been predicted by the approximate
analytical description of an unstable mass-transfer evolution
presented by \citet{WebIben}.

The information plotted in Figure \ref{Q1.3-D J plots} allows us to
examine, in part, the role that ``direct impact'' accretion plays in
driving the evolution of this particular binary system. Throughout
most of the mass-transfer phase of the evolution, the accretor
gradually spins up (the dashed curve labelled $J_a$ in the bottom
panel) as material from the stream is deposited onto its surface at
an oblique impact angle.  It is significant that the curves for
$J_a(t)$ and $J_\mathrm{orb}(t)$ are practically mirror images of
one another, while the curve for $J_d(t)$ is almost perfectly flat
until the final ``plunge'' occurs.  This means that the spin-up of
the accretor occurs almost entirely at the expense of the orbit.
That is, at the time of impact, the specific angular momentum of the
accretion stream material does not reflect the specific angular
momentum of the star from which it originated ({\it i.e.}, the
donor) but instead reflects the amount by which the time-varying
torque applied by the system's complex (and time-varying)
gravitational field has been able to transfer angular momentum from
the orbit to the stream as the stream material ``falls'' from the
vicinity of the $L_1$ Lagrange point to the surface of the accretor.
This result provides strong support for the arguments that have led
\citet{Maet04} and \citet{GPF} to account for the effects of
direct-impact accretion in their semi-analytical models of
mass-transferring binary systems through a term that is built around
the relatively simple concept of a circularization radius, as
described above in the context of Eq.~(\ref{adot}). At the very end
of the ``mass-transfer'' phase of the Q1.3-D evolution, the binary
loses orbital angular momentum to the spin of both stellar
components catastrophically, as the stars plunge towards one another
and eventually merge.

The above discussion suggests the following interpretation:
initially and during most of the evolution, while the angular
momentum of the donor remains flat, the system is evolving as a
result of the mass transfer instability since $q>q_{\rm stable}$.
Only near the end, tidal effects on the donor further reduce the
orbital angular momentum causing the final merger. Thus this final
phase should be considered the proper tidal instability. Binary
systems in which direct impact occurs will have a harder time than
disk systems avoiding the tidal instability since $q_{\rm
stable}<2/3$.

\subsubsection{Evolution Q1.3-E}

In an attempt to ascertain to what extent the outcome of a
mass-transfer event depends on the manner in which the donor star is
brought into contact with its critical Roche surface, we performed a
second simulation in which the initial state was defined by model
Q1.3 (with the associated parameters shown in Table
\ref{table1_1.3}) but, instead of artificially extracting angular
momentum from the system, during the ``driving'' phase of the
evolution we gradually increased the specific entropy of the gas
inside the donor star.  This caused a slow, secular increase in the
effective radius of the donor and, within the first two orbits,
brought the surface of the donor into sufficiently deep contact with
its critical Roche surface that the mass-transfer event was
initiated.  More specifically, for this Q1.3-E evolution the
polytropic constant $K_d$ for the material inside the donor was
increased at a rate of 1.67\% of its initial value per orbit for 2
orbits.  (See Appendix~\ref{drivingAppendix} for a precise description of how this
``driving'' technique was implemented in the hydrocode.) Because
$R\propto K$ in an $n=3/2$ polytropic star (see, for example, the
discussion in \S\ref{secBackground} associated with
Eq.~\ref{PolytropicM_R}), this level of driving should have caused
the effective radius of the donor to increase by approximately 3.3\%
by the end of the driving phase of the Q1.3-E evolution.

As was anticipated, the results of this Q1.3-E simulation were very
similar to the results of the Q1.3-D simulation.  The accretion
stream that had become well-defined by the end of the ``driving''
phase grew steadily thicker throughout the ``mass-transfer'' phase
and, shortly after the system mass-ratio dropped below unity, the
cores of the two stars catastrophically merged into a single object.
There were, however, quantifiable differences between the two model
evolutions.  Most noticeably, the stars in the Q1.3-E evolution
merged at an earlier evolutionary time.  We expected that the two
evolutions would be somewhat offset in time from one another.  Given
that two distinctly different mechanisms were employed to drive the
system into contact, it was unlikely that, for example, $\dot{M}_d$
would be precisely the same in both simulations at the end of the
driving phase (beginning of the mass-transfer phase) hence, it was
unlikely that a basic system parameter such as $q$ would be
precisely the same at a given time, $t$, because its value depends
on the {\it integral} over $\dot{M}_d$ up to the time $t$. Before
making further comparisons between these two simulations, we decided
to synchronize their evolutionary times at the end, rather than at
the beginning, of the mass-transfer phase. More specifically, we
took a plot of $\log{|\dot{M}_d|}$ versus time from the Q1.3-D
simulation (solid curve in the top panel of Figure \ref{Q1.3-D-E Mqa
plots}) and slid it horizontally across the analogous plot from the
Q1.3-E simulation --- normalized in the same manner as the top panel
of Figure \ref{Q1.3-D-E Mqa plots} ---  until a reasonable match
between the two time-dependent functions was achieved at late times.
In this manner we were able to identify a ``merger time'' for the
Q1.3-E evolution that could be reasonably well associated with the
merger time $t_\mathrm{merge}$ that was defined earlier for the
Q1.3-D evolution. For the Q1.3-E evolution, a time offset of $5.9
P_0$ was required to align the origin in time with the end of the
mass-transfer phase, that is, $t_\mathrm{merge}\equiv (t_E -
5.9P_0)$, where $t_E$ is the evolutionary time recorded during the
Q1.3-E simulation. By comparison, a time offset of $12.0 P_0$ was
required to arrange the same alignment for the Q1.3-D evolution.

The dotted curves in the top, middle, and bottom panels of Figure
\ref{Q1.3-D-E Mqa plots} display the behavior of, respectively,
$\log(|\dot{M}_d|/\dot M_{\rm ref})$, $q$, and $a/a_0$ as a function
of $t_\mathrm{merge}/P_0$ from simulation Q1.3-E.  The mass-transfer
rate and the function $q(t_\mathrm{merge})$ for this simulation
match the mass-transfer rate and the function $q(t_\mathrm{merge})$
for simulation Q1.3-D very well over the entire ``mass-transfer''
phase of its evolution. The dotted curve for
$\dot{M}_d(t_\mathrm{merge})$ exhibits lower-amplitude oscillations
than the corresponding solid curve, indicating that the binary orbit
has remained more nearly circular throughout the Q1.3-E evolution.
In the plot of the time-dependent behavior of the binary separation
(bottom panel of Figure \ref{Q1.3-D-E Mqa plots}), the solid curve
lies below the dotted curve at all times.  This is understandable
because, due to the manner in which the system was initially driven
into contact, the model in simulation Q1.3-D (solid curve) has a
slightly smaller orbital angular momentum than the model in
simulation Q1.3-E (dotted curve) at all times.

The top panel of Figure \ref{Q1.3-D-E Mqa plots} explicitly shows
that, at the end of the driving phase of simulation Q1.3-E ($t_E =
2P_0$, hence $t_\mathrm{merge} = - 3.9 P_0$), the mass-transfer rate
was higher than it was at the end of the driving phase of simulation
Q1.3-D ($t_D = 2 P_0$, hence $t_\mathrm{merge} = -10 P_0$).  As was
forecast, above, this in itself explains why the Q1.3-E simulation
merged more quickly. Had we stopped ``driving'' the expansion of the
donor somewhat sooner --- say, after only 1.5 orbits instead of
after 2 orbits
--- $\dot{M}_d$ would have been smaller at the onset of the
mass-transfer phase of simulation Q1.3-E and it would have taken
longer for the two stars to merge. Nevertheless, the mass-transfer
phases of the two separate Q1.3 model simulations are remarkably
similar and certainly the outcome of the evolutions (catastrophic
merger) is the same.  This supports our expectation that the
endpoint of a mass-transfer evolution will be insensitive to the
manner in which the binary system is initially brought into contact.

\begin{figure}[!ht]
\centering \includegraphics[scale=0.8,viewport=67 573 521 719,clip]{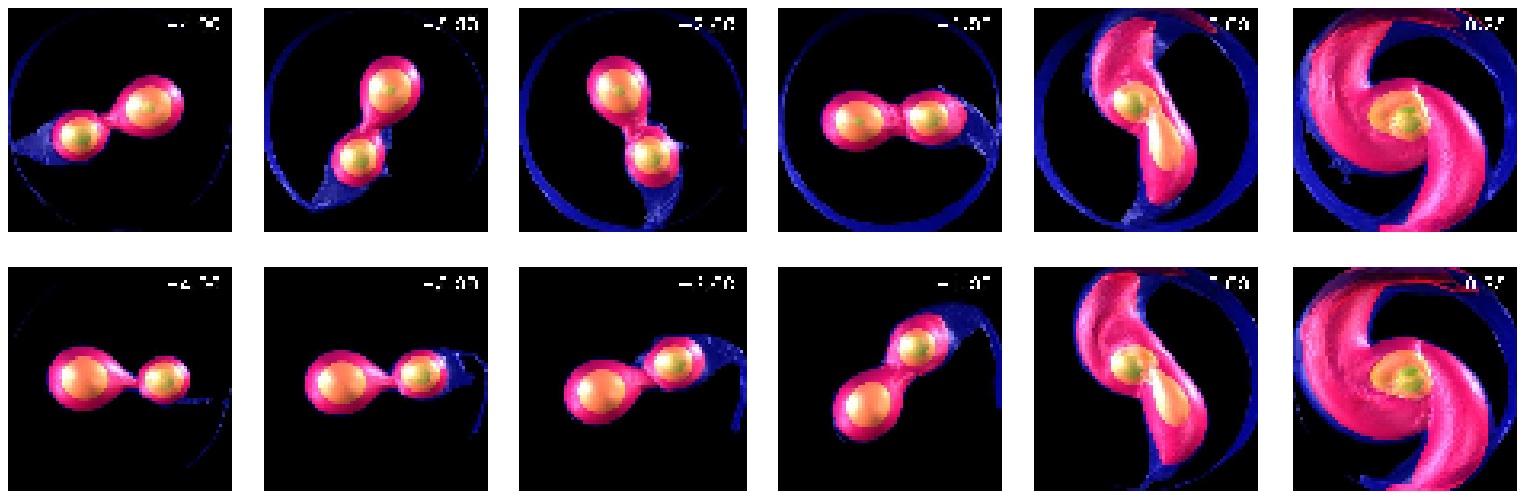} \caption{Two evolutions
with $q_0=1.3$, in which mass transfer is initiated by angular
momentum losses (Q1.3-D, top row) and by expansion of the donor
(Q1.3-E, bottom row), lead to an almost identical merger. The entire
Q1.3-E evolution is depicted in a mpeg animation in the online
version of this article. The time shown in the movie is $t_{\rm
E}$.} \label{Q1.3-D-E images compared}
\end{figure}
The top half of Figure \ref{Q1.3-D-E images compared} contains six
frames from the animation sequence (mentioned earlier) that was
generated to visually illustrate the complex fluid-dynamical flows
that developed during simulation Q1.3-D.  Four of these images are
identical to ones that were presented earlier in Figure \ref{Q1.3-D
images}, but here the evolutionary time $t_\mathrm{merge}$ has been
used to identify when each image was drawn from the simulation. For
comparison, images from simulation Q1.3-E are displayed in the
bottom half of Figure~\ref{Q1.3-D-E images compared}  at the same
synchronized times, $t_\mathrm{merge}$. Each of the last two images
in the bottom row of Figure~\ref{Q1.3-D-E images compared} is
remarkably similar to its corresponding image in the top row of the
Figure, given that they represent highly distorted structures that
have developed from complex, nonlinear, mass-transfer evolutions and
given that they have arisen at two quite different ``absolute''
times in the separate numerical simulations.  (At earlier times,
{\it e.g.}, $t_\mathrm{merge} = - 1 P_0$, the binary structures are
also quite similar, although there is an identifiable orbital phase
discrepancy due to the fact that the orbital separations and, hence,
the orbital frequencies are slightly different in the two
simulations.)  The correspondence between these highly distorted
structures at late times in these two separate simulations provides
perhaps the strongest empirical evidence that the endpoint of a
mass-transfer evolution will be insensitive to the manner in which a
binary system is initially brought into contact. At the same time,
the quantitative similarity between the late-time results of
evolutions Q1.3-D and Q1.3-E serves as a strong, affirmative
convergence test for our hydrocode.

\subsection{Simulations with $M_{\rm d} = 0.5 M_{\rm
a}$ Initially}\label{sec_0.5_results}

\begin{figure}[!ht]
\centering \includegraphics[scale=0.8,viewport=67 498 521 719,clip]{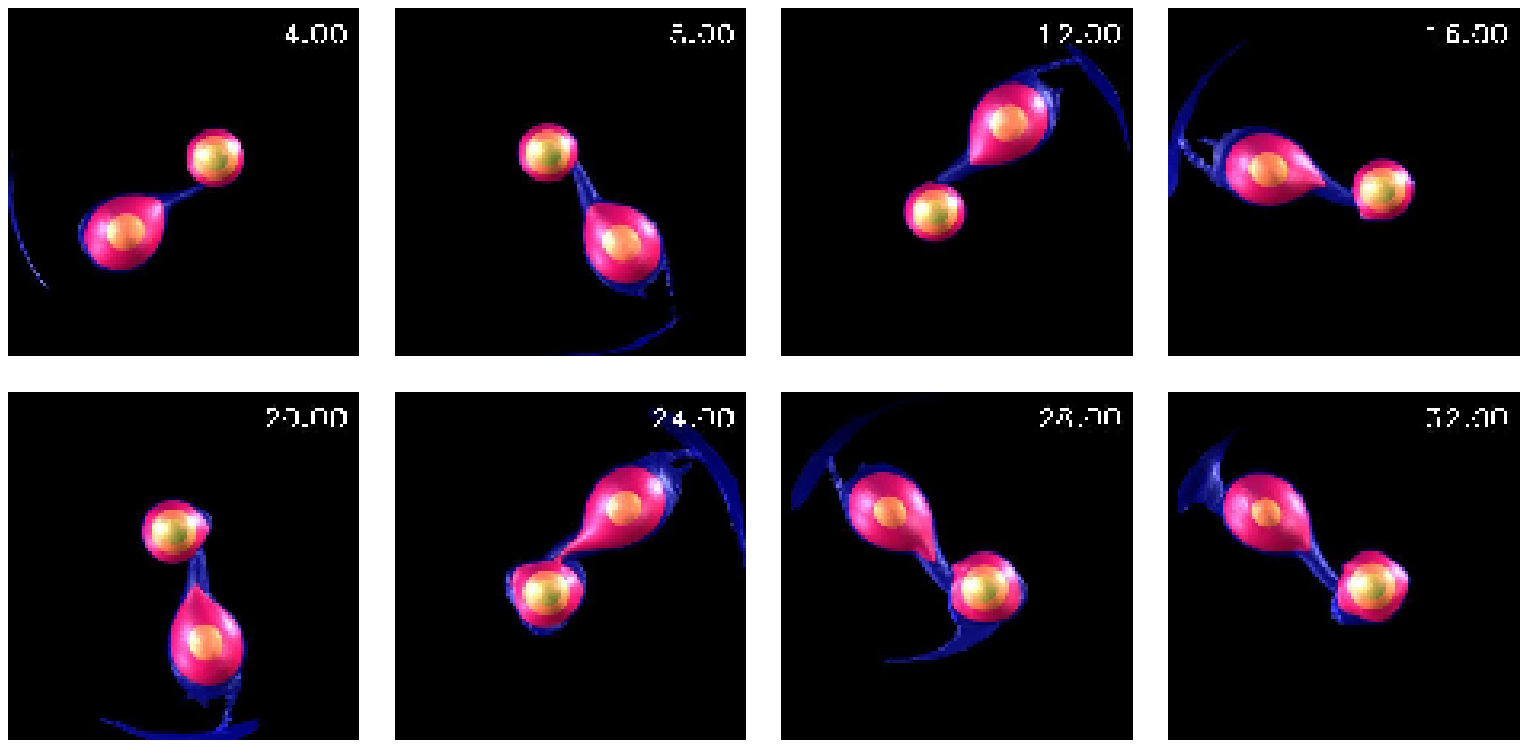} \caption{Same as Figure
\ref{Q1.3-D images}, but for our Q0.5-Da evolution; images are
separated in time by four initial orbital periods. Even after 32
orbits, the system continues to undergo mass transfer and there is
no indication that the system will merge. The entire evolution is
depicted in the mpeg animation accompanying the online article.}
\label{Q0.5a images}
\end{figure}

\begin{figure}[!ht]
\centering \includegraphics[scale=0.8,viewport=12 16 465 592,clip]{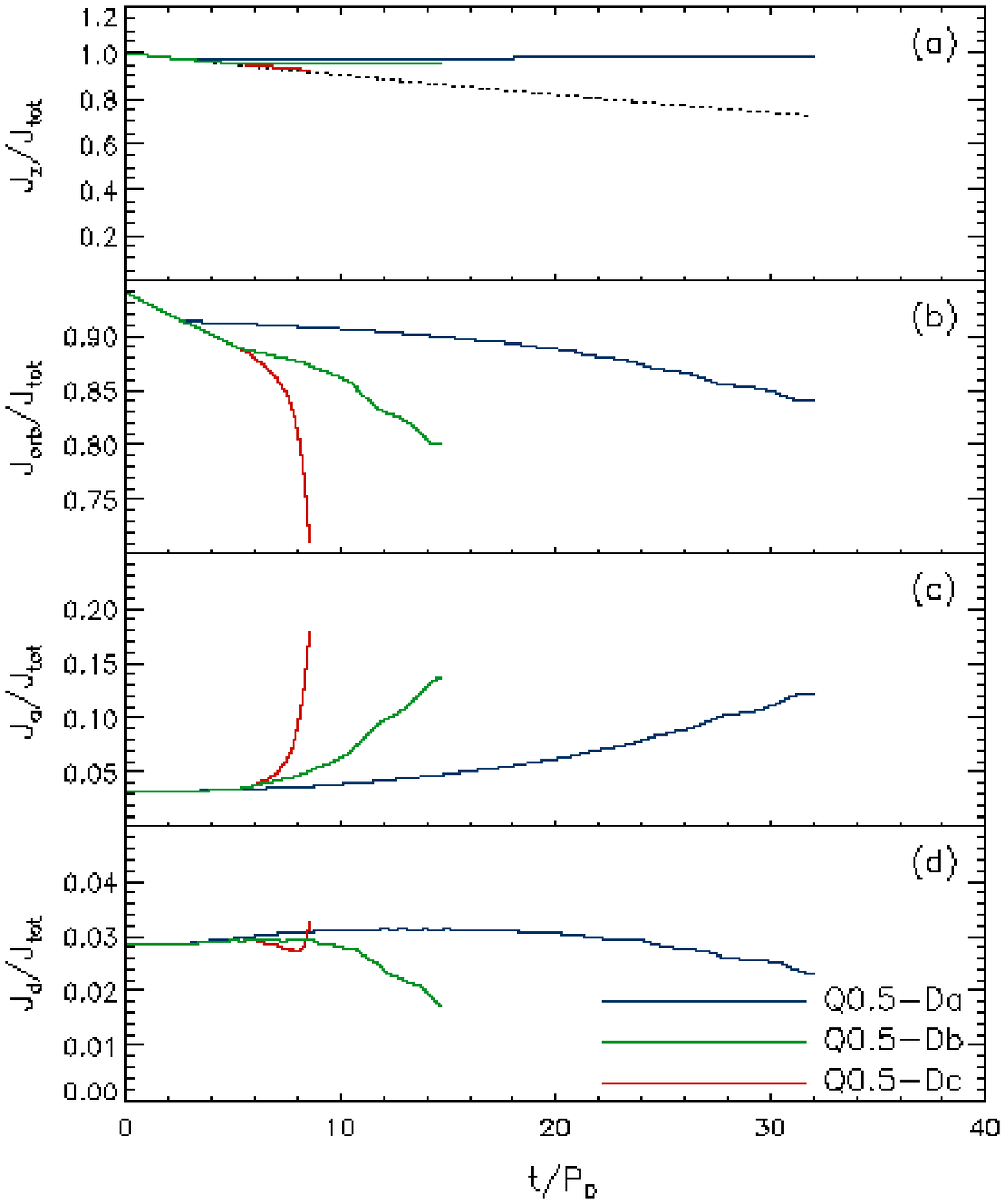} \caption{The same
quantities as in Figure \ref{Q1.3-D J plots}, but for the Q0.5-D
evolutions: Q0.5-Da (blue), Q0.5-Db (green), and Q0.5-Dc (red). Note
that the vertical scale for panel (d) showing $J_{\rm d}/J_{\rm
tot}$ has been expanded for clarity.} \label{Q0.5 J plots}
\end{figure}

\begin{figure}[!ht]
\centering \includegraphics[scale=0.8,viewport=12 16 465 592,clip]{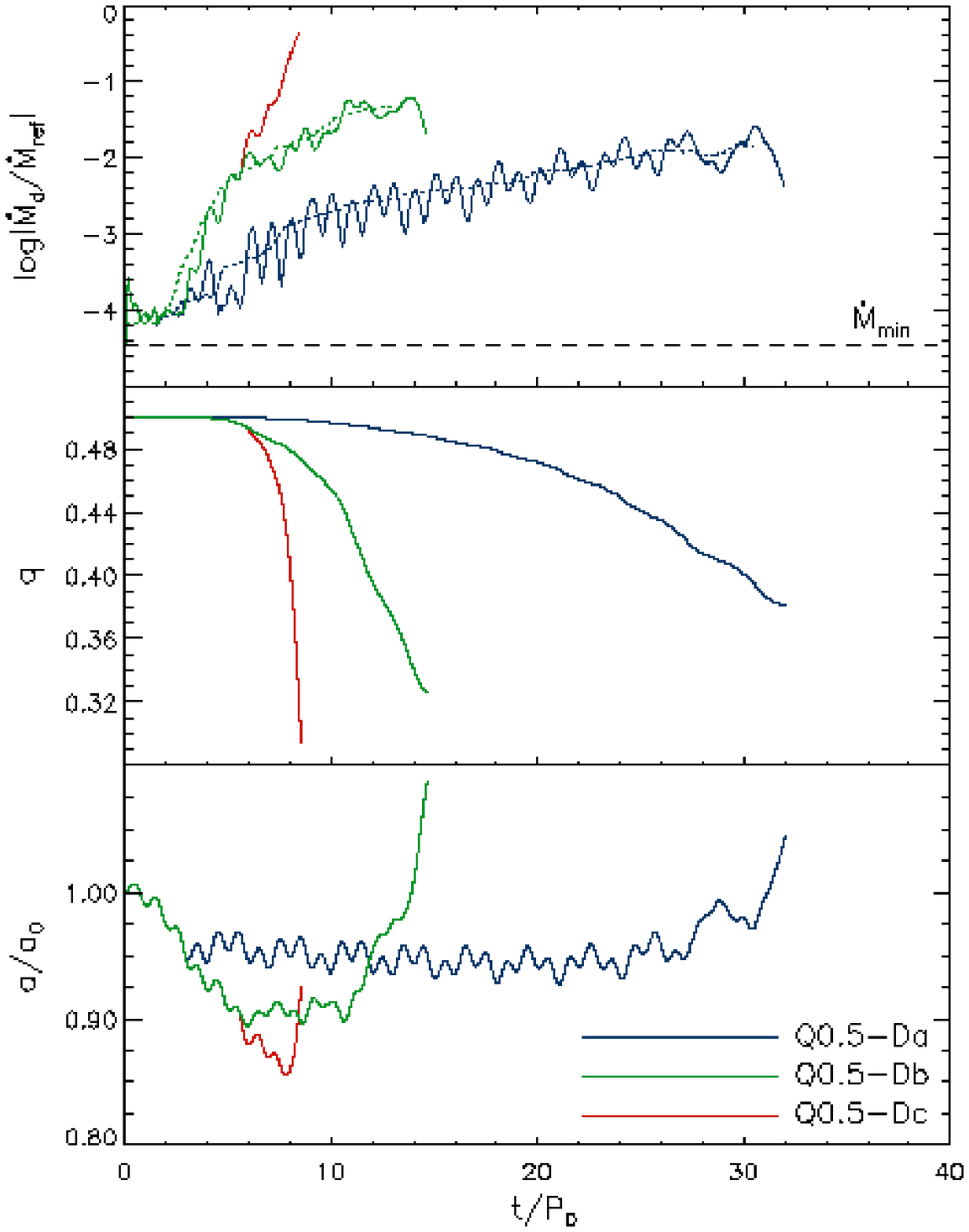} \caption{Same as Figure
\ref{Q1.3-D-E Mqa plots} but for the Q0.5-D evolutions: Q0.5-Da
(blue), Q0.5-Db (green), and Q0.5-Dc (red). All quantities are
plotted versus the evolutionary time measured in units of the
initial orbital period of the system. For evolutions Q0.5-Da and
Q0.5-Db we also show in the top panel, as dashed lines, a 3-orbit
boxcar average of the mass transfer rate.} \label{Q0.5 Mqa plots}
\end{figure}

\begin{figure}[!ht]
\centering \includegraphics[scale=0.8,viewport=12 480 600 700,clip]{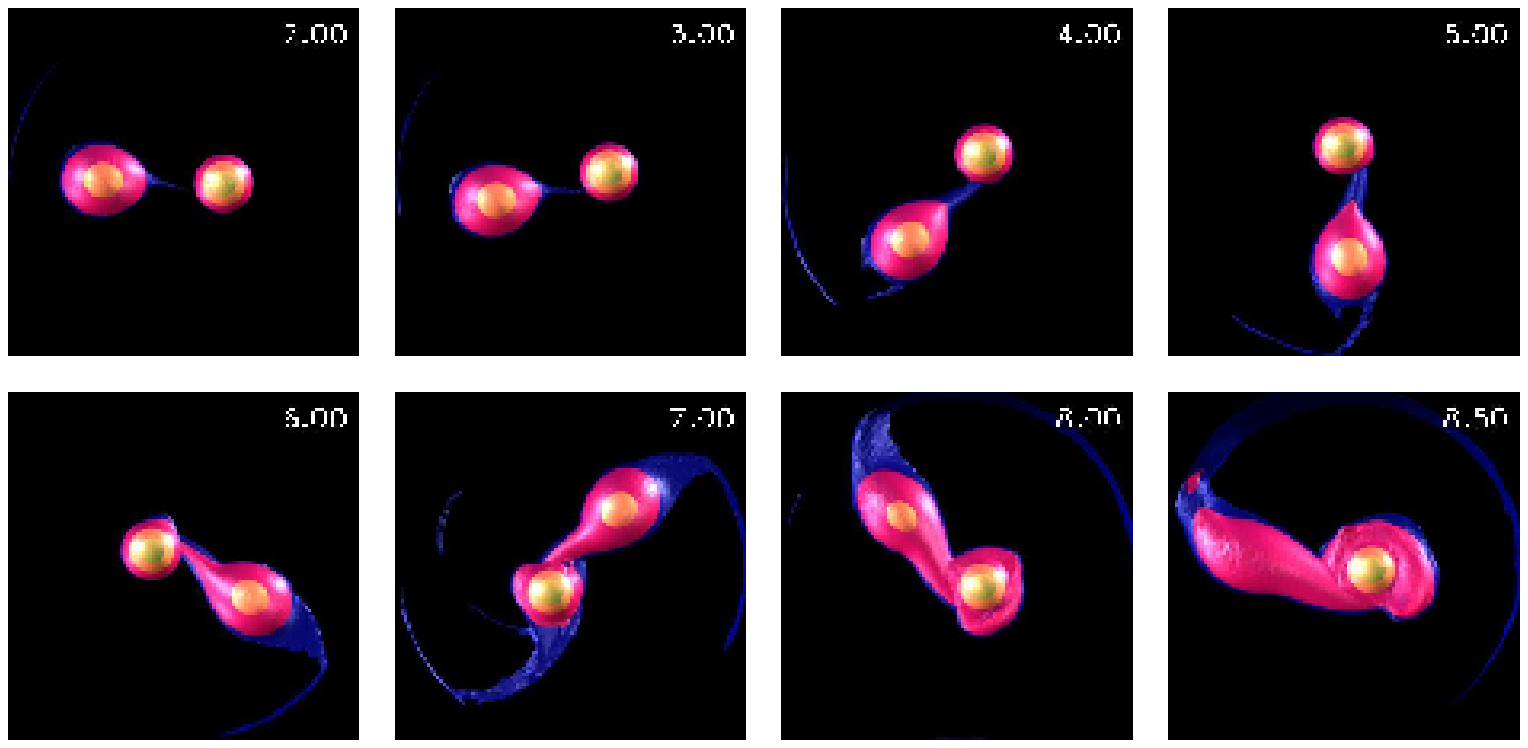} \caption{Same as Figure
\ref{Q0.5a images}, but for our Q0.5-Dc evolution; images are
separated in time by one initial orbital period, except the last
image. Since this binary is continuously driven, the separation and
the orbital period decrease throughout, except perhaps at the very
end.} \label{Q0.5c images}
\end{figure}

As an example of mass transfer in binaries with $q<2/3$, we now
discuss the results of three Q0.5-D simulations in which driving
was applied for different times and compare the results with those
obtained by RS95 for a DWD binary of identical mass ratio
(hereafter, referred to as simulation Q0.5-RS). Building on our
detailed analysis and description of the model Q1.3 evolutions,
the Q0.5-D simulations can be described fairly concisely. In
particular, Figures \ref{Q0.5a images}-\ref{Q0.5c images} display
selected results from our Q0.5-D evolutions in the same manner in
which results from simulation Q1.3-D were presented in Figures
\ref{Q1.3-D images}-\ref{Q1.3-D-E Mqa plots}.

\begin{deluxetable}{lcclcc}
\tabletypesize{\scriptsize}

\tablecaption{Initial Parameters for Model Q0.5\tablenotemark{\dag}} \tablewidth{0pt}

\tablehead{ \colhead{System} & \colhead{Initial} & ~ &
\colhead{Component} &
~ & ~ \\
\colhead{Parameter} & \colhead{SCF Value} & ~ & \colhead{Parameter}
& \colhead{Donor} & \colhead{Accretor}  }

\startdata
$q_0$ & 0.500 & ~~~~~~~~~~ & $M_i$ & $3.073\times 10^{-3}$ & $6.143\times 10^{-3}$
\\
$a_0$ & 0.8764 & ~~~~~~~~~~ & $\rho_i^\mathrm{max}$ & 0.235 & 1.0000   \\
$\Omega_0$ & 0.1174 & ~~~~~~~~~~ & $K_i$ & 0.016 & 0.016   \\
$J_\mathrm{tot}$ & $1.97\times 10^{-4}$ & ~~~~~~~~~~ & $V_i$ & 0.0814 & 0.0370   \\
${R}_\mathrm{com}$ & $1.04\times 10^{-5}$ & ~~~~~~~~~~ & $V_i^\mathrm{RL}$ & 0.0906
& 0.2380   \\
\enddata
\tablenotetext{\dag}{Parameter values are given here in dimensionless
polytropic units. To scale these numbers to other (e.g., cgs) units, see the
discussion in Appendix~\ref{unitsAppendix}.}
\label{table0.5}
\end{deluxetable}

Initially, the donor star in our model Q0.5-D simulations was
slightly detached from its Roche lobe: $V_{\rm d}/V^{\mathrm{\rm
RL}}_{\mathrm{d}} = 0.90$ (Table \ref{table0.5}).  In the first of
our three Q0.5-D runs, henceforth simulation Q0.5-Da, in order to
bring the system into contact and initiate the mass-transfer event,
angular momentum was drained from the binary
--- the ``driving'' phase of the simulation --- at a rate of 1\% per
orbital period over 2.7 periods. The system was then allowed to
evolve in the absence of driving
--- the ``mass-transfer'' phase of the evolution --- since there
was no driving present in the simulations of RS95 and because our
primary objective was to see if mass transfer, once initiated, is
unstable. In a second Q0.5-D simulation, subsequently referred to
as Q0.5-Db, we extended the driving phase to 5.3 orbits at the
same level of 1\% per orbit, and then let it evolve without
driving. This allowed us to investigate the effects of deeper
initial contact and a higher mass-transfer rate on the evolution
of the binary. Finally, in a third simulation, referred to as
Q0.5-Dc, we applied driving at the rate of 1\% per orbit
throughout the evolution. The length of time over which angular
momentum was extracted from the system in all three of these
Q0.5-D simulations is reflected in panel (a) of Figure~\ref{Q0.5 J
plots}. During the driving phase the angular momentum follows
closely the dotted line, turning horizontal when the driving stops
for evolutions Q0.5-Da (blue curve) and Q0.5-Db (green curve). The
Q0.5-Dc evolution (red curve) is driven throughout, so it follows
the dotted line to the end. According to expectations, the longer
we applied driving, the shorter the time it took for the binary to
evolve to a comparable mass transfer level and evolutionary stage.
Consequently, the longest of these three simulations was Q0.5-Da
(32 orbits), while Q0.5-Dc was the shortest (8.5 orbits).

\begin{figure}[!ht]
\centering \includegraphics[scale=0.8,viewport=12 16 465 592,clip]{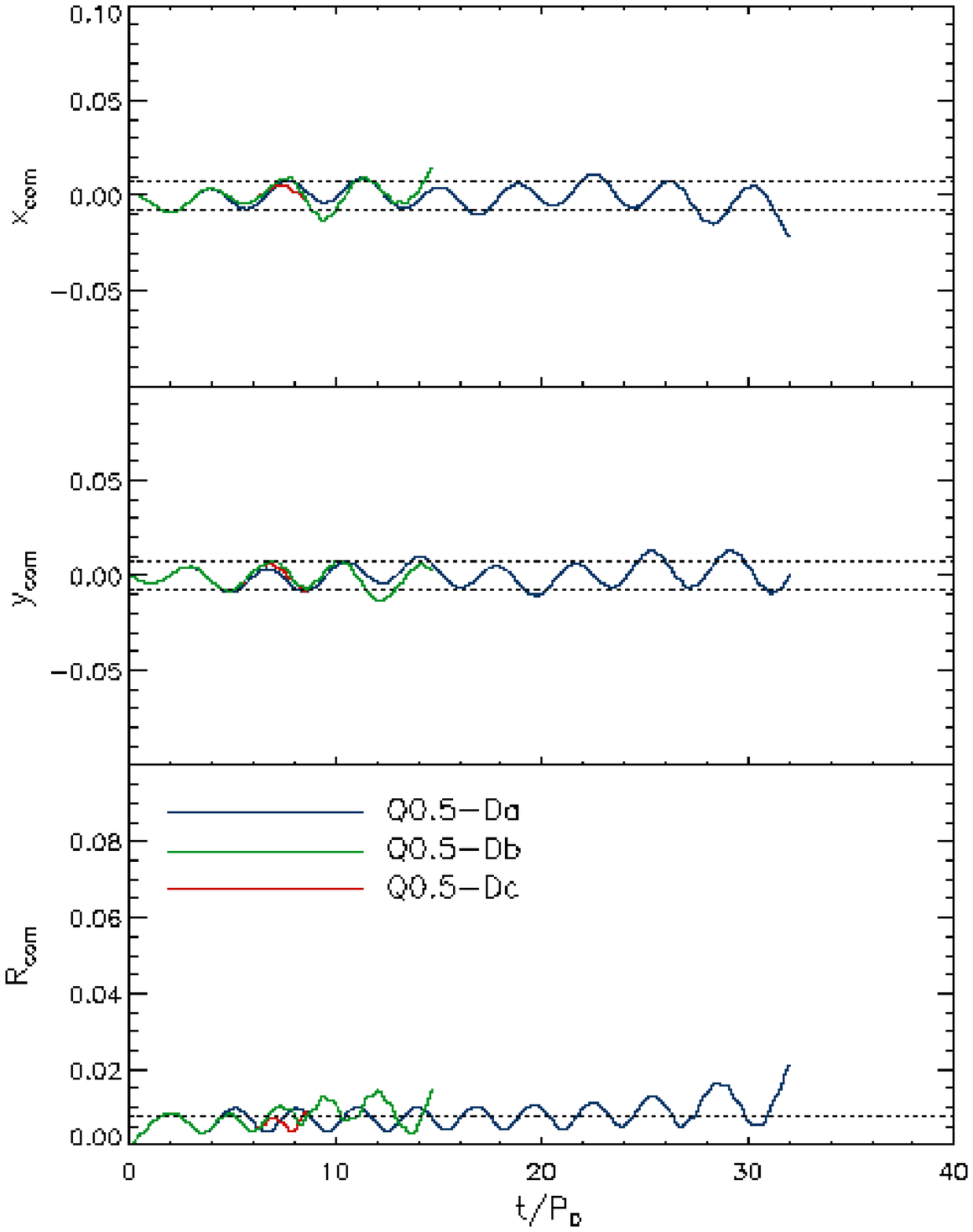} \caption{Drift of the
center of mass during the Q0.5-D simulations: Q0.5-Da (blue),
Q0.5-Db (green), and Q0.5-Dc (red). From top to bottom: the $x$ and
$y$ components of the position of the center of mass in the
corotating frame, and the cylindrical radial distance to the center
of mass $R_{\rm com}$, as functions of $t/P_0$. The dotted lines at
$\pm \Delta R$ show the extent of the innermost radial grid zone.
The center of mass stays very nearly within this one radial grid
zone throughout these evolutions.} \label{Q0.5 xyRcom}
\end{figure}

We focus first on Figure \ref{Q0.5a images}, which displays images
of the mass-density distribution at selected time slices during
the Q0.5-Da evolution, and on the blue curve in the bottom panel
of Figure \ref{Q0.5 Mqa plots}, which shows the time-evolutionary
behavior of the binary separation for this same model evolution.
During the brief ``driving'' phase of the evolution, the
separation decreases at a rate that is consistent with the rate at
which angular momentum is being extracted from the system.  An
epicyclic oscillation with a period $\approx P_0$ and an amplitude
of just over 1\% is present in the $a(t)$ curve; as before, we
suspect this arises from initial conditions and driving.   In
addition there is another oscillation with a period of about
$3P_0$ which appears to be related to a slow meandering of the
position of the center of mass of the system (see the discussion,
below, related to Figure \ref{Q0.5 xyRcom}). We note that these
oscillations also appear --- although at a somewhat reduced
amplitude --- in evolutions Q0.5-Db (green curve) and Q0.5Dc (red
curve), and they are mirrored in the mass transfer rates shown in
the top panel of Figure \ref{Q0.5 Mqa plots}.

After driving has been turned off in evolution Q0.5-Da, the
separation hovers around a value $a \approx 0.95 a_0$ for more
than twenty orbits, then the binary begins to separate and the
mass-transfer rate (blue curve in the top panel of Figure
\ref{Q0.5 Mqa plots}) levels off.  The images in Figure \ref{Q0.5a
images} reflect this $a(t)$ behavior. Initially the binary
advances in the corotating frame as its orbital frequency
increases. At late times, as the separation increases, the binary
slows down, stalls and after some hesitation due to the various
oscillations described above, it moves in a retrograde sense when
the orbital frequency falls below $\Omega_0$ at $t\approx 31 P_0$.
(This behavior is quite clear in the $\approx 3800$-frame
animation sequence from which the individual images shown here
were extracted.)  In summary, we have followed the Q0.5-Da
evolution through more than 30 orbits in the presence of a steady
mass-transfer stream and there is no indication that the system is
going to merge.  This result is significantly different from the
Q0.5-RS evolution, which was violently unstable to mass transfer
and led to tidal disruption of the donor within $\sim 5$ orbits.

In an effort to understand why our Q0.5-Da evolution differed from
the Q0.5-RS evolution, we extended the initial ``driving'' phase to
$5.3$ orbits (model Q0.5-Db) in order to bring the Roche lobe into
deeper contact with the donor and bring the system to a higher
mass-transfer rate as it entered the ``mass-transfer'' phase of its
evolution. As the green curve in the bottom panel of Figure
\ref{Q0.5 Mqa plots} shows, the orbital separation steadily
decreased (as expected) during the extended ``driving'' phase of
this evolutions, but after the driving ceased the separation $a(t)$
evolved in a slightly different manner from the Q0.5-Da evolution.
Specifically, the system did not hover at its minimum separation as
long; it began separating and its mass-transfer rate began to level
off after only $\approx 11 P_0$. We can understand this difference
in behaviors if we refer to Eq.~(\ref{adot}).

In the absence of tidal effects, Eq.~(\ref{adot}) predicts that
the orbital separation should increase as soon as driving is
interrupted. [Note that this result holds even when there is a
significant consequential transfer of orbital angular momentum to
the spin of the accretor --- see panels (b) and (c) of Figure
\ref{Q0.5 J plots} where, as was seen in the Q1.3-D evolution,
$J_\mathrm{a}(t)$ is practically a mirror image of
$J_\mathrm{orb}(t)$.]  However, after the driving phase has ended
in evolution Q0.5-Da, the separation does not immediately begin to
increase; in fact, a careful examination of the blue curve in the
bottom panel of Figure \ref{Q0.5 Mqa plots} shows that the
separation continues to decrease, albeit at a much reduced rate.
This deviation from the ``expected'' behavior must be attributed
to the tidal terms in Eq.~(\ref{adot}).
The blue curve in panel (d) of Figure \ref{Q0.5 J plots} provides
evidence that tides are at work:  over the first approximately 14
orbits of evolution Q0.5-Da, the donor is also being spun up,
storing part of the orbital angular momentum and thereby allowing
the separation to continue to decrease slowly.  This is readily
understandable because the initial driving rapidly shrinks the
binary, thereby increasing the orbital frequency. The spin of the
donor lags initially and recovers gradually as it is spun up by
tides. Some time after $J_{\rm d}$ peaks at $t\sim 14 P_0$, the
binary separation finally begins to increase at $t\sim 18-20 P_0$,
and the mass-transfer rate levels off. In evolution Q0.5-Db, the
separation stops decreasing almost immediately after the driving
phase has ended (see the green curve in Figure \ref{Q0.5 Mqa
plots}). It increases slowly at first, then much more rapidly at
the end of the simulation. Because the driving phase lasted longer
in this evolution than in Q0.5-Da, the mass-transfer rate was able
to climb to a sufficiently high level by the time driving was
stopped to permit the ``$\dot{M}_d$'' term in Eq.~(\ref{adot}) to
exceed the negative tidal terms.

At the end of both the Q0.5-Da and Q0.5-Db simulations, not only is
the binary separation increasing, but the mass-transfer rate has
leveled off. We speculate that if we were able to follow these
evolutions significantly farther in time in the absence of driving,
the magnitude of $\dot{M}_d$ would steadily decrease and we would
find that the binary eventually detaches and mass-transfer ceases.
We also conjecture that the mass of the remnant donor would be a
decreasing function of the level and duration of the original phase
of driving. Confirmation of these conjectures must await further
improvements in our simulation tools (see further discussion,
below).

In order to investigate how an even higher mass-transfer rate
might affect the evolution of this $q_0 = 0.5$ binary system, we
performed a third simulation (model Q0.5-Dc) with continuous
``driving'' at a rate of 1\% per orbit.  Because driving was never
turned off, systemic angular momentum losses played a dominant
role in dictating how the orbital parameters of the system varied
throughout most of this evolution. As the red curve in the bottom
panel of Figure \ref{Q0.5 Mqa plots} shows, the orbital separation
steadily decreased for $\approx 8 P_0$ at a rate that would be
predicted by the first term alone on the right-hand-side of
Eq.~(\ref{adot}). As a consequence, the Roche lobe sank quite deep
into the envelope of the donor and, as depicted by the red curve
in the top panel of Figure \ref{Q0.5 Mqa plots}, the mass-transfer
rate steadily grew throughout the evolution, reaching a level that
was an order of magnitude higher than the maximum rate acquired in
evolution Q0.5-Db and roughly two orders of magnitude higher than
the maximum rate acquired in evolution Q0.5-Da.  Even under these
extreme conditions, however, the orbital separation eventually
reached a minimum (at $t/P_0 \approx 8$) and the system began
separating (red curve in the bottom panel of Figure \ref{Q0.5 Mqa
plots}).  Presumably this reversal occurred because the
mass-transfer rate became large enough for the ``$\dot{M}_d$''
term in Eq.~(\ref{adot}) to finally dominate over systemic angular
momentum losses.

As the images in Figure \ref{Q0.5c images} illustrate, through
approximately $7 P_0$, model Q0.5-Dc evolved through configurations
that resemble those seen in the Q0.5-Da evolution (see Figure
\ref{Q0.5a images}).  But late in the evolution, the accreted
material has formed a much more prominent, time-dependent,
nonaxisymmetric disk-like structure around the accretor; and at the
end of the simulation, the donor is being tidally ripped apart, even
as the measured orbital separation is increasing. This evolution was
terminated when a significant fraction of material from the tidally
elongated donor hit the outer edge of the computational domain.
Interestingly, the evolution of this ``driven'' system during its
final 2-3 orbits bears a strong resemblance to the final 2-3 orbits
of the published Q0.5-RS evolution.

For evolutions Q0.5-Da and Q0.5-Db, we think that our results are
more aptly described as the donor being gradually stripped of its
mass and partially disrupted by tides rather than as a
catastrophic merger of the binary. We suspect that, if these
evolutions could be followed accurately beyond what the present
version of our code is able to do, a sizeable portion of the donor
would survive (final mass ratio $q\sim 0.3$) in an elliptical
orbit. In fact, even in evolution Q0.5-Dc it appears that although
the donor is largely tidally disrupted, a small remnant may
survive at a larger radius. Unfortunately, because of the growing
drift of the center of mass (see Figure~\ref{Q0.5 xyRcom}), we
cannot follow evolutions Q0.5-Da and Q0.5-Db far enough to
unambiguously show that a long-lived remnant of the donor survives
the tidal disruption. Despite these limitations, our results are
significantly different from the Q0.5-RS evolution, which was
violently unstable to mass transfer and led to tidal disruption of
the donor within $\sim 5 $ orbits. As was briefly forecast in
\S\ref{SecExpectations}, and as has been discussed in early
paragraphs of this section of the paper, the results we have
obtained are consistent with the behavior expected from
Eq.~(\ref{adot}). They also are consistent with the results of
integrations of orbit-averaged evolution equations to be described
elsewhere \citep{GPF}.

A possible reason that the Q0.5-RS binary was found by RS95 to be
unstable to mass transfer is that the donor may have been in deeper
contact with its critical Roche surface at the start of their
simulation. To test this we have calculated the values of the
separation $r$ (as defined in RS95) at the point the driving is
terminated in our simulations and have compared it with RS95. In
case Q0.5-Da, $r\approx 4.0$, which is greater than the initial
separation, $r = 3.9$, reported by RS95 for evolution Q0.5-RS (note
that in RS95 the donor is in contact with its Roche lobe at the
outset of the simulation). It appears as though the two stars are
initially closer to each other in simulation Q0.5-RS and, hence, the
donor is in deeper contact with its critical Roche surface
initially. In turn, this implies that the mass-transfer simulation
conducted by RS95 started with a higher accretion rate. From Figure
13a of RS95, $\log{\vert \dot{M}_{\rm d}\vert}\approx -1$ after only
4 orbits, which is higher than the mass transfer rate we observed at
any stage of our Q0.5-Da simulation. In case Q0.5-Db, $r\approx
3.8$, very close to the initial conditions of RS95, and yet the
accretion rate appears to peak below $\log{\vert \dot{M}_{\rm
d}\vert}\approx -1$. Finally, in run Q0.5-Dc, the separation appears
to turn around at the end when $r\approx 3.6$, above the value at
which we expect the tidal instability to set in. Unfortunately,
because the donor bumps up against the edge of our computational
domain at the end of our simulation, we cannot follow model Q0.5-Dc
far enough to deduce its ultimate fate. However, the accretion rate
is still growing rapidly while the orbital angular momentum is
plummeting suggesting a full tidal disruption and perhaps eventual
merger. Determining whether the donor survives in this case remains
an obvious goal for future simulations.

We also note that in the Q0.5-RS evolution, the distance between the
centers of mass of the two stars decreases while the density maxima
of the two components separate. This behavior seems odd, but can
perhaps be understood if the mass in the stream is large enough to
significantly shift the center of mass of the distorted donor in the
downstream direction. In our Q0.5-D simulations the mass in the
stream is always a small fraction of the donor's mass. As a result,
both the distance between the centers of mass of the binary
components and the separation of their density maxima increase at
late times thus saving the donor from tidal disruption, except
perhaps in the Q0.5-Dc case.

It should be emphasized that, in our Q0.5-D simulations, the
adopted initial driving rate was orders of magnitude larger than
what one would expect in a realistic DWD binary. A milder driving
favors stability since all effects discussed above will also be
milder. With driving applied throughout the entire Q0.5-Dc evolution
at the initial rate of 1\% per orbit, a tidal disruption may well be
the final outcome, and it remains to be seen if any fraction of the
donor survives. The behavior observed in this case comes closest
qualitatively to the evolution reported by RS95 for the Q0.5-RS
simulation.

Finally, Figure \ref{Q0.5 xyRcom} shows the time-dependent
meandering of the position of the center of mass of the Q0.5 binary
system throughout our Q0.5-D simulations.  Its radial distance from
the cylindrical coordinate axis, $R_\mathrm{com}$ (bottom panel),
and its associated equatorial-plane Cartesian coordinates
($x_\mathrm{com}$, top panel; and $y_\mathrm{com}$, middle panel) as
viewed from a frame of reference rotating with the initial orbital
frequency, $\Omega_0$, are plotted as a function of $t/P_0$. As was
described by MTF and discussed in \S \ref{commotion} above, there is
a tendency for an unequal-mass binary system to drift away from the
cylindrical coordinate axis during a simulation. If left unchecked,
some part of the binary would hit the boundary of the grid in a time
$\sim 10 P_0$. During our Q0.5-D simulations the corrections
described in \S \ref{commotion} have succeeded in confining the
center of mass to within $\sim 1$ zone of the computational grid up
to the time when the binary begins to separate rapidly and the donor
approaches the boundary. In run Q0.5-Da the simulation remains
well-behaved for over 30 orbits. This is significantly longer than
any other self-consistent hydrodynamic simulation of a binary
evolution with or without mass-transfer that we are aware of.

Despite this success, however, there is still room for improvement.
For example, it appears that the residual center-of-mass motion
shown in Figure \ref{Q0.5 xyRcom} has been reflected in an
undesirable way in other dynamical features of simulation Q0.5-D.
Most noticeably, the oscillation that is seen in all three panels of
Figure \ref{Q0.5 xyRcom} with a period $\sim 3 P_0$ and an amplitude
$\sim \Delta R/a_0 = 0.9\%$ appears to be modulating the natural
epicyclic oscillations that appear in the functions $a(t)$ and
$\dot{M}_{\rm d}(t)$ throughout the mass-transfer phase of the
evolution (see the bottom and top panels of Figure \ref{Q0.5 Mqa
plots}). In addition, we have observed that, with the correction in
place, as the binary begins to separate rapidly, a numerical
instability can eventually cause the center of mass to rapidly
spiral outward.  We have stopped the Q0.5-D simulations before this
instability sets in.

\section{Discussion and Conclusions}
\label{Conclusions}

This is the second in a series of papers that describe results of
direct simulations of the dynamical evolution of unequal-mass
binaries using a three-dimensional, finite-difference hydrodynamics
technique.  Our long-term goal is to gain a better understanding of
the origin and survival of various classes of binaries --- including
double white dwarfs, contact binaries and direct-impact accretors
--- by accurately simulating hydrodynamical flows that arise when
binaries of various kinds become semi-detached. In contrast to this,
much of the related analytic and numerical work that has been
carried out over the past decade has placed an emphasis on the onset
of the tidal instability and on the coalescence and merger of nearly
identical neutron stars. In these earlier studies, mass transfer
events were modeled when necessary as a prelude to the almost
inevitable merger.  In fact, the inevitability of mergers seems to
have been accepted also for double white dwarf binaries.  While this
may well be the case with massive white dwarfs, the results
presented here already suggest that low-mass WD binaries may escape
merger and survive as AM CVn systems \citep{TY96, NPVY01}.

In the present paper, after describing improvements that have been
made in the hydrodynamics code that was developed by and described
in MTF, we have taken the first steps toward elucidating the role
that is played by mass transfer in the outcome of the dynamical
phase of evolution following first contact. We have been especially
concerned with those binary systems in which the mass ratios and
equations of state are such that Roche lobe contact occurs before a
tidal instability, so we have focused on binaries with $n=3/2$
($\gamma=5/3$) polytropic components \citep{RS95, UE1, UE2}. We have
presented two evolutions of a system having an initial mass ratio
$q_0 = 1.323$ that was dynamically unstable to mass transfer. In one
evolution, mass transfer was initiated by removing angular momentum
from the binary (Q1.3-D) while in the other evolution we expanded
the donor star causing it to overflow its Roche lobe (Q1.3-E). While
there were subtle differences in the structure of the binary and the
depth of contact by the time driving was terminated, in both cases
the binary was dynamically unstable and a merger was the final
result. In fact, we demonstrated that if one looks at the last two
or three orbital periods before the merger, the behavior of both
evolutions is consistent with being the same. We therefore conclude
that, for unstable systems, it does not matter how the system gets
into contact; once the instability kicks in, the behavior is the
same. This result also has served as a convergence test for our
hydrodynamics code since two evolutions with identical initial
states but that were brought into contact in different ways and
therefore had different detailed histories, ended up looking
indistinguishable during the final stages leading to a merger.

At the very end of the ``mass-transfer'' phase of both the Q1.3-D
and Q1.3-E evolutions, the binary transferred orbital angular
momentum to the spin of both components catastrophically, as the
stars plunged towards one another and eventually merged.  Overall
the behavior we observed during the final few orbits agrees with the
predictions of the tidal instability \citep{LRS1, LRS2, LRS3, LRS4,
LRS5}. This suggests that one can recognize two epochs during the
evolution of a $q_0 > q_{\rm stable}$ binary system that undergoes
an episode of mass-transfer: Over an extended period of time, the
rate of mass transfer grows while the orbital angular momentum
changes very slowly. This instability is driven by an ever
increasing depth of contact and thus can be considered a
mass-transfer instability. Once the separation has been reduced
sufficiently, the second instability sets in, during which the
orbital angular momentum drops rapidly while the spin angular
momenta increase driven by tides.

We have also presented detailed results from three mass-transfer
evolutions (simulations Q0.5-Da, Db, and Dc) of a polytropic binary
with an initial mass ratio $q_0=0.5$ and in which the two components
had the same specific entropy. Thus this initial state could
represent a DWD system with components having the same composition.
It also corresponds to one system that was simulated by \citet{RS95}
using an SPH technique. In each of our simulations, the less massive
star was driven into contact with its Roche lobe via the slow
removal of orbital angular momentum, but in simulations Q0.5-Da and
Q0.5-Db this artificial driving was turned off shortly after the
mass-transfer event was initiated (after 2.7 and 5.3 initial orbital
periods, respectively). Our results differ in some important ways
from the results reported by RS95.  Most significantly, in the two
simulations that were evolved for an extended period of time in the
absence of systemic angular momentum losses, we do not observe a
merger nor a tidal disruption; in evolution Q0.5-Da (Db) our binary
survives for more than 30 (14) orbital periods and at the end it is
separating while the mass-transfer rate has leveled off. Via these
extended simulations, we have demonstrated that our numerical tools
permit us to accurately model accretion flows in dynamically
evolving mass-transfer systems.  We can, for example, analyze how
angular momentum is exchanged between the orbit and the spin of the
two stars during a phase of direct-impact accretion, and we can
analyze how the structure of the accretor dynamically readjusts as
relatively high specific angular momentum material is deposited onto
its equatorial region. Studies of this type should assist in the
determination of which simplifying assumptions are justified -- as
well as which are not -- in models that attempt to describe extended
phases of mass-transfer evolutions semi-analytically \citep{WebIben,
Maet04}. By conducting a variety of related simulations, we
ultimately hope to be able to determine what the critical mass ratio
$q_\mathrm{stable}$ is that defines which binary systems are stable
or unstable against mass-transfer.

We speculate that the outcomes of our Q0.5-Da and Q0.5-Db
simulations were different from the Q0.5-RS simulation because the
evolution presented in \citet{RS95} was started from a significantly
deeper initial contact, and thus transferred a larger fraction of
the mass before separating. In RS95, the maximum density of the
donor appears to be moving away and yet the separation calculated as
the distance between the centers of mass of the material in their
respective Roche lobes seems to be decreasing. Therefore, our
results could be described as reproducing some of the initial
features of the SPH simulation by \citet{RS95} at a slower pace.
Evidently, if the mass ratio of the binary is such that initially
the system is unstable to mass transfer, the final fate of the
binary will depend on whether during the initial phase of mass
transfer the separation decreases sufficiently for the tidal
instability to take over. In our Q0.5-Dc simulation, in which
driving remained on throughout the evolution, it appears as though
the system encounters the tidal instability; the last 2-3 orbits of
this evolution resemble fairly closely the evolutionary behavior of
the published Q0.5-RS simulation.  However, in our Q0.5-Da and
Q0.5-Db simulations, once the driving is cut off the tidal effects
only succeed in delaying the tendency of the binary to separate as
mass transfer proceeds by temporarily storing some orbital angular
momentum in the spin of the donor. Once the separation begins to
increase, tides become ineffective and the binary avoids the merger.

Given an equation of state for the binary components, the mode of
mass transfer, and the expected mass loss, if any, it is possible
to make predictions about the evolutionary outcome after contact.
However, our simulations suggest that the eventual fate of such a
binary depends not only on the initial mass ratio $q_0$, but also
on how far $q_0$ is above the appropriate $q_{\rm stable}$, and
even on the rate of driving unless this is very slow. In other
words, the detailed outcome depends on the non-linear development
of the mass-transfer and tidal instabilities. Only when $q_0$ is
well above $q_{\rm stable}$ will the evolution of mass transfer
proceed rapidly enough to resemble qualitatively the predictions
of the analytic solution derived in \citet{WebIben}. When $q_0$ is
only slightly above $q_{\rm stable}$, non-linear effects come into
play that make it possible for the system to survive the
mass-transfer instability and avoid merger. We shall discuss these
questions further in two forthcoming papers \citep{GPF, MDTF}.
As has already been mentioned, the tools we have developed will
also enable us in the future to investigate in more detail the
hydrodynamics of mass transfer and the structures arising from
this transfer, transient flows, oscillations, mixing and
convection. We see already some of these features in our
simulations and we think they warrant further investigation.

\acknowledgements This work has been supported in part by NSF grants
AST 04-07070 and PHY 03-26311, and in part through NASA's ATP
program grants NAG5-8497 and NAG5-13430.  The computations were
performed primarily at NCSA through grant MCA98N043, which allocated
resources on the Tungsten cluster, and on the SuperMike and
SuperHelix clusters at LSU, which are operated by the Center for
Computation and Technology (CCT).  J. E. T. acknowledges support
from the NSF-sponsored Institute for Pure and Applied Mathematics at
UCLA, which provided an environment in May, 2005 that was conducive
to writing significant portions of this manuscript. We thank the
referee for many insightful comments and for encouraging us to
perform some additional simulations.

\appendix

\section{Correction for Center-of-Mass Motion}\label{comAppendix}

In \S\ref{commotion}, we presented an analytic expression for a
small ``artificial'' acceleration ${\bf a}^\mathrm{art}$ that has
been added to the source term of the equation of motion in an effort
to counteract a slow wandering of the center-of-mass of the binary
system in each of our simulations. Here we describe how the various
terms in expression (\ref{comAcceleration}) have been evaluated in
the hydrocode.

In practice, at each discrete point in time $t^{(n)}$, the location
${\bf r}_\mathrm{com}^{(n)} = [{\bf i}x^{(n)}_\mathrm{com} + {\bf
j}y^{(n)}_\mathrm{com} + {\bf k}z^{(n)}_\mathrm{com}]$ of the center
of mass of the system has been determined numerically via an
appropriate integral over the mass-density distribution,
$\rho^{(n)}$. Then, for example, knowing the $x$-component of the
center-of-mass position from the present and two previous points in
time, $t^{(n-1)}$ and $t^{(n-2)}$, the $x$-components of the
center-of-mass velocity and acceleration have been determined
empirically through the respective finite-difference expressions,
\begin{eqnarray}
\dot{x}_{\mathrm{com}}^{(n)} &=& \frac{x^{(n)}_\mathrm{com}\{2\Delta
t^{(n-1)}\Delta t^{(n-2)} + [\Delta t^{(n-2)}]^2\} -
x^{(n-1)}_\mathrm{com}[\Delta t^{(n-1)} + \Delta t^{(n-2)}]^2 +
x^{(n-2)}_\mathrm{com}[\Delta t^{(n-1)}]^2}{\Delta t^{(n-1)}\Delta
t^{(n-2)}[\Delta
t^{(n-1)}+\Delta t^{(n-2)}]} \, , \\
\ddot{x}_{\mathrm{com}}^{(n)} &=&
2\biggl\{\frac{x^{(n)}_\mathrm{com} [\Delta t^{(n-2)}] -
x^{(n-1)}_\mathrm{com} [\Delta t^{(n-1)} + \Delta t^{(n-2)}] +
x^{(n-2)}_\mathrm{com} [\Delta t^{(n-1)}]}{\Delta t^{(n-1)}\Delta
t^{(n-2)}[\Delta t^{(n-1)}+\Delta t^{(n-2)}]}\biggr\} \, ,
\end{eqnarray}
where $\Delta t^{(n-1)} \equiv [t^{(n)} - t^{(n-1)}]$ and $\Delta
t^{(n-2)} \equiv [t^{(n-1)} - t^{(n-2)}]$.  Both of these
expressions were derived from a straightforward Taylor-series
expansion between the discrete coordinate positions at various
times; analogous expressions produced measured values of the other
cartesian components of the center-of-mass velocity and
acceleration, $\dot{y}_{\mathrm{com}}^{(n)},
\dot{z}_{\mathrm{com}}^{(n)}, \ddot{y}_{\mathrm{com}}^{(n)},$ and
$\ddot{z}_{\mathrm{com}}^{(n)}$. Finally, expressed in terms of the
cylindrical coordinates used in the hydrocode,
\begin{eqnarray}
a_{R,\mathrm{com}}^{(n)} = \ddot{x}_{\mathrm{com}}^{(n)}\cos\phi +
\ddot{y}_{\mathrm{com}}^{(n)}\sin\phi\, &;&~~
v_{R,\mathrm{com}}^{(n)} = \dot{x}_{\mathrm{com}}^{(n)}\cos\phi +
\dot{y}_{\mathrm{com}}^{(n)}\sin\phi \,
, \\
a_{\phi,\mathrm{com}}^{(n)} = \ddot{y}_{\mathrm{com}}^{(n)}\cos\phi
- \ddot{x}_{\mathrm{com}}^{(n)}\sin\phi\, &;&~~
v_{\phi,\mathrm{com}}^{(n)} = \dot{y}_{\mathrm{com}}^{(n)}\cos\phi -
\dot{x}_{\mathrm{com}}^{(n)}\sin\phi \,
, \\
a_{z,\mathrm{com}}^{(n)} = \ddot{z}_{\mathrm{com}}^{(n)} \, &;& ~~
R_{\mathrm{com}}^{(n)} = \{[{x}_{\mathrm{com}}^{(n)}]^2
+[{y}_{\mathrm{com}}^{(n)}]^2\}^{1/2} \, . \label{RcomDefined}
\end{eqnarray}

\section{Conversion of Code Units to Physical Units}
\label{unitsAppendix}

All of the initial model parameter values in Tables
\ref{table1_0.843}, \ref{table1_1.3}, and \ref{table0.5} are given
in units such that $G = R_\mathrm{SCF} =\rho^\mathrm{max}_{\rm
a}(t=0) = 1$, where $R_\mathrm{SCF}$ is the outer edge of the
cylindrical grid that was used to generate the model in the SCF
code.  Fairly straightforward relationships can be used to scale
the values of these dimensionless parameters to more meaningful
({\it e.g.}, cgs) units.  For example, if the total mass and
orbital period of the Q0.8 binary configuration are specified in
cgs units as $M_\mathrm{cgs}$ and $P_\mathrm{cgs}$, respectively,
then the orbital separation of the system in centimeters is,
\begin{equation}
a_\mathrm{cgs} =  \biggl[ \frac{\Omega_0^2 a_0^3}{M_{\rm a} +
M_{\rm d}} \biggr]^{1/3}
\biggl[\frac{GMP^2}{4\pi^2}\biggr]^{1/3}_\mathrm{cgs} = 1.015~
\biggl[\frac{GMP^2}{4\pi^2}\biggr]^{1/3}_\mathrm{cgs} \,
\end{equation}
the effective radius of the accretor in centimeters is,
\begin{equation}
R_a^\mathrm{cgs} = \frac{a_\mathrm{cgs}}{a_0} \biggl(
\frac{3V_{\rm a}}{4\pi} \biggr)^{1/3} =  \biggl[ \frac{3V_a
\Omega_0^2}{4\pi(M_a+M_d)}
\biggr]^{1/3}\biggl[\frac{GMP^2}{4\pi^2}\biggr]^{1/3}_\mathrm{cgs}
= 0.351~\biggl[\frac{GMP^2}{4\pi^2}\biggr]^{1/3}_\mathrm{cgs} \, ,
\end{equation}
and the maximum density of the donor is,
\begin{equation}
\rho_d^\mathrm{cgs} = \rho^\mathrm{max}_d \biggl[
\frac{M_\mathrm{cgs}}{M_{\rm d} + M_{\rm a}} \biggr]
\biggl(\frac{a_\mathrm{cgs}}{a_0}\biggr)^{-3} =
\frac{\rho_d^\mathrm{max}}{\Omega_0^2} \biggl[ \frac{4\pi^2}{P^2
G} \biggr]_\mathrm{cgs} = 68.1~\biggl[ \frac{3\pi}{P^2 G}
\biggr]_\mathrm{cgs} \, .
\end{equation}
If, instead, the configuration's reduced mass $\mu_\mathrm{cgs} =
M_\mathrm{cgs}[q_0/(1+q_0)^2]$ and orbital frequency
$\Omega_\mathrm{cgs} = 2\pi/P_\mathrm{cgs}$ are specified in cgs
units, then the system's total angular momentum in cgs units is,
\begin{equation}
J_\mathrm{cgs} = J_\mathrm{tot} \biggl[
\frac{\Omega_0}{(M_a+M_d)^5} \biggr]^{1/3} \biggl[
\frac{(1+q_0)^2}{q_0} \biggr]^{5/3} \biggl[
\frac{G^2\mu^5}{\Omega} \biggr]^{1/3}_\mathrm{cgs} = 2.82 ~\biggl[
\frac{G^2\mu^5}{\Omega} \biggr]^{1/3}_\mathrm{cgs} \, .
\end{equation}
The leading numerical coefficient in the final expression for each
one of these relations has been obtained by plugging in values of
$\Omega_0$, $a_0$, $V_a$, $M_d$, etc. drawn from Table
\ref{table1_0.843}.  The numerical value of each of these leading
coefficients must be reevaluated using the dimensionless parameter
values drawn from Table \ref{table1_1.3} or Table \ref{table0.5}
if one wants to scale our Q1.3 or Q0.5 models, respectively, to
cgs units.

Note that no mass-radius relationship, or equivalently value for the
polytropic constant $K_\mathrm{cgs}$, has been assumed above, and thus
the polytropic binary model can represent arbitrary types of component stars.
One is free to choose any total mass and orbital period, and in most cases
the resultant binary model will not represent any realistic type of
donor or accretor. However, choosing an orbital period consistent with
the mass-radius relationship of a given type of star will yield approximate
binary models with components of that type. For example, a Q0.8 model
with a total mass of $1.29 M_\odot$, can represent equally well a main sequence binary
with an orbital period of $\approx 5.4$ hours, or a white dwarf binary
with a period of $\approx 70$ seconds.

\section{Implementation of Driving Mechanisms}
\label{drivingAppendix}

In the initial Q1.3 and Q0.5 models (see Figure 1 and Tables 4 and
5) the donor star slightly underfilled its Roche lobe and had to be
driven into contact to initiate mass transfer. As has been outlined
in the introductory paragraphs of \S 5, we tested two driving
mechanisms to initiate Roche lobe overflow. Here we describe
specifically how these were implemented in the hydrodynamic code.

\subsection{Drag}
During the first 2-3 orbits of evolutions Q1.3-D and Q0.5-D we
extracted angular momentum from the binary at a constant rate. This
was accomplished by adding a ``sink'' term ($s_{\rm{drag}}$) to the
azimuthal component of the equation of motion. Specifically, the
right-hand-side of Eq. (15) was modified to give,
\begin{eqnarray}\label{azim_eqn}
\frac{\partial A}{\partial t} + \mbox{\boldmath$\nabla$} \cdot (A
\mbox{\boldmath$\upsilon$} ) &=& - \frac{\partial p}{\partial \phi}
- \rho \frac{\partial \Phi}{\partial \phi} - 2 \Omega_0 S R - \rho
R[a_{\theta,\mathrm{com}} +2\Omega_0 v_{R,\mathrm{com}}] -
s_{\rm{drag}} \, ,\
\end{eqnarray}
where $s_{\rm{drag}} \equiv + \alpha_{\rm{drag}} [\Omega_{0}
R^{2}/P_0] \rho$ and $\alpha_{\rm{drag}}$ is a constant that fixes
the rate at which specific angular momentum is removed. In
evolutions Q1.3-D and Q0.5-D we set $\alpha_{\rm{drag}}$ = 0.01.

For the record, this sink term was incorporated in the hydrodynamics
code by effectively adding to the right-hand-side of MTF's Eq. (42)
a term of the form,
\begin{eqnarray}
- \alpha_{\rm{drag}} [\Omega_0 R^{2}_{i+1/2}/P_0]
\hat{\rho}^{(n+\rm{advection)}}_{i+1/2,j+1/2,k}\, ,\
\end{eqnarray}
where the superscript, subscript, and caret notations are as defined
in MTF.

\subsection{Expansion}
During the first 2 orbits of evolution Q1.3-E, the specific entropy
of the material that was originally in the donor was steadily
increased in order to drive a slow expansion of the donor.  This was
accomplished by adding a source term ($s_{\rm expand}$) to Eq. (17)
to give,

\begin{eqnarray}
\frac{\partial \tau}{\partial t} + \mbox{\boldmath$\nabla$} \cdot
(\tau \mbox{\boldmath$\upsilon$} ) &=& s_{\rm{expand}} \, ,
\end{eqnarray}
where $s_\mathrm{expand} \equiv \alpha_\mathrm{expand}\tau/P_0$ and
the expansion coefficient was set to $\alpha_\mathrm{expand} =
0.01$.  Specifically in the hydrodynamics code, after the updates
due to the Eulerian transport terms were performed as described by
Eq.~(39) of MTF, the entropy tracer was further updated according to
the finite-difference expression,
\begin{eqnarray}
\tau_{i+1/2,j+1/2,k+1/2}^{(n + \rm{source})} =
\tau_{i+1/2,j+1/2,k+1/2}^{(n+\rm{advection})}\biggl[1 +
\alpha_\mathrm{expand}\frac{\Delta t}{P_0} \biggr] \, ,
\end{eqnarray}
where the superscript and subscript notations are as defined in MTF.
Utilizing the hydrocode's ability to track which fluid originally
belonged to which star, this update to $\tau$ was applied only to
those computational cells in which the fraction of the donor
material in that cell was greater than 90\% of the total value.

In order to estimate how a given change in $\tau$ effects the radius
of the donor, we note first that for a polytropic gas of index
$n=3/2$,
\begin{eqnarray}
\tau = \rho \biggl[\frac{3}{2}K_\mathrm{d}\biggr]^{3/5} \, .
\end{eqnarray}
Hence, increasing $\tau$ by a factor of $[1+\alpha_\mathrm{expand}
\Delta t/P_0]$ effectively increases $K_\mathrm{d}$ by a factor of
$[1+(5/3)\alpha_\mathrm{expand} \Delta t/P_0]$.  Furthermore, from
the discussion associated with Eq.~(3) we see that, for a given
mass, the radius of a spherical polytrope of index $3/2$ scales
linearly with $K$.  Setting $\alpha_\mathrm{expand} = 0.01$
therefore should result in the radius of the donor expanding at a
rate of 1.67\% per orbit.


\end{document}